\documentclass[letterpaper,12pt]{article}
\usepackage{tabularx} 
\usepackage{amsmath}  
\usepackage{amssymb,braket,bm}
\usepackage{graphicx}
\usepackage{subcaption}
\usepackage[margin=1in,letterpaper]{geometry} 
\usepackage{cite} 
\usepackage[normalem]{ulem}
\usepackage{tikz}

\usepackage[final]{hyperref} 
\hypersetup{
	colorlinks=true,       
	linkcolor=blue,        
	citecolor=blue,        
	filecolor=magenta,     
	urlcolor=blue         
}
\usepackage{authblk}

\begin{document}

\title{$\mathbb{Z}_2$ topologically ordered phases \\
on a simple hyperbolic lattice
}

\author{Hiromi Ebisu$^{1,2}$, Bo Han$^1$}
\affil{$^1$Department of condensed matter Physics, Weizmann institute of science, Rehovot, Israel}
\affil{$^2$Department of Physics and Astronomy, Rutgers, The State University of New Jersey,
School of Arts and Sciences, 
New Jersey, USA}
\date{\today}
\maketitle

\begin{abstract}
In this work, we consider 2D $\mathbb{Z}_2$ topologically ordered phases ($\mathbb{Z}_2$ toric code and 
the modified surface code) on a simple hyperbolic lattice.  
Introducing a 2D lattice consisting of the product of a 1D Cayley tree and a 1D conventional lattice, we investigate two topological quantities of the $\mathbb{Z}_2$ topologically ordered phases on this lattice: the ground state degeneracy on a closed surface and the topological entanglement entropy.
We find that both quantities  
depend on the number of branches
and the generation of the Cayley tree. We attribute these results to a huge number of superselection sectors of anyons. 
\end{abstract}
\maketitle
 \section{Introduction}
Topologically ordered phases are novel phases of matter beyond the paradigm of the standard Ginzburg–Landau theory~\cite{tsui,wen1989chiral,wen1990,Kitaev2003}. There are many salient features in these phases, such as non-trivial ground state degeneracy (GSD)
when the systems are placed on manifolds with non-trivial topology~\cite{Elitzur:1989nr}, and fractionalized quasi-particle excitations (anyons)~\cite{leinaas1977theory,Wilczek1982,laughlin1983anomalous}. 
Topologically ordered phases have spurred a great deal of interest, involving different branches of physics. Examples are topological quantum field theories~\cite{1982Jackiw,witten1989quantum}, quantum error correcting codes~\cite{dennis2002topological}, universal quantum computations~\cite{Nayak2008}, and the classification of symmetry protected 
topological phases by investigating anyonic statistics after gauging global symmetry~\cite{dijkgraaf1990topological,spt2013,PhysRevB.86.115109,Kapustin2014}.


\par
While topologically ordered phases are intensively studied on Euclidean lattices, less is well understood when they are placed on the \textit{non-Euclidean lattices}.  
The motivation of this work is to study $\mathbb{Z}_2$ topologically ordered phases on a simple \textit{hyperbolic lattices}, which are negatively curved manifolds, and explore the interplay between topologically ordered phases and geometric structures of the lattice.
Specifically, we focus on 
one simplest example of such lattices, the Cayley tree\footnote{One can intuitively understand this by recalling the fast that there exists a tessellation of the hyperbolic plane constructed from $n$-gons with $k$-polygons meeting at each vertex if $\frac{1}{n}+\frac{1}{k}<\frac{1}{2}$, 
and taking the limit $n\to\infty$, which gives rise to
the Cayley tree with $k$-branches ($k>2$). }, which has been intensively studied in the context of statistical mechanics (see, for instance,  Ref.~\cite{fisher1961some}). \par
In order to investigate 
how topological properties of the $\mathbb{Z}_2$ topologically ordered phases on the Cayley tree (Fig.~\ref{z2}) are different from those on the conventional Euclidean lattice, 
we focus on two aspects of the topological properties. The first point is to construct closed surface and count the GSD.  The crucial properties of topologically ordered phases is that when the theory is placed on a non-trivial manifold, non-contractible loops of anyons give rise to non-trivial GSD.
The second point is to investigate entanglement entropy in a bipartite system separated by a cylindrical geometry. In particular, we analyze 
the topological entanglement entropy~\footnote{To be more precise, we study the non-local entanglement entropy proposed in~\cite{kim2014,PhysRevB.97.144106}.}, which depends only on universal contributions~\cite{levinwen2006,preskillkitaev2006}. 
The hallmark of the lattice being hyperbolic is that the wavefunction of an anyon is delocalized, giving rise to a huge number of superselection sectors (i.e., the number of distinct types of anyons), depending on the coordination number $k$ and the generation $M$ of the Cayley tree. We find that such dependence can be clearly seen in both of the GSD and the topological entanglement entropy. 

Relations between topologically ordered phases and the hyperbolic lattice have been mainly discussed in the context of quantum error corrections, trying to find an efficient code (see, e.g., Refs.~\cite{freedman2002z2,zemor2009cayley}). There are several works to elucidate the properties of fracton topological phases, which are new type of phases of matter beyond the conventional topologically ordered phases, on hyperbolic geometry in recent years~\cite{manoj2021arboreal,Han2019}. In addition, hyperbolic geometries have been used to study band structure, quantum computing and holography in different areas of physics~\cite{Breuckmann2017hyperboliccode,maciejko2021hyperbolic,PastawskiYoshidaHarlowPreskill2015holographiccode}. The novelty in our paper compared with the previous works is that we propose a systematic way to study the superselection sectors and the corresponding anyon excitations in the hyperbolic spacetime. To be more specific, we study the unusual fusions rules and the $S$ and $T$ matrices characterizing the statistics of the anyons. We also explore how these excitations contribute to the nonlocal entanglement entropy that is well-studied in the planar geometry. We believe there exist close connections between our work and the Bloch wavefunction and the Berry phase that are formulated in hyperbolic band theories~\cite{maciejko2021hyperbolic}. We hope our model will provide an interesting playground for further research.  \par 
The outline of this paper is as follows. In Sec.~\ref{m22}, we introduce the lattice and Hamiltonian. In Sec.~\ref{sec tor}, we construct a closed surface of the lattice and study GSD. We also present two types of matrices characterizing the statistics of excitations and stability argument against local perturbations.  Sec.~\ref{nonlocal enta} is devoted for calculating the entanglement entropy of the system and see how topological entanglement entropy characterizes the interplay between quasi-particle excitations and geometry of the Cayley tree. We further give a consistency check by imposing boundary terms in the model, corroborating our quasi-particle interpretations on the entanglement entropy. Finally, in Sec.~\ref{6}, we give conclusion and discuss future directions. Details of study on other $\mathbb{Z}_2$ topological phases as well as 
technical details are relegated to appendices.



\section{Model}\label{m22}

\begin{figure}
  \begin{center}
   \begin{subfigure}[h]{0.14\textwidth}
       \includegraphics[width=1.4\textwidth]{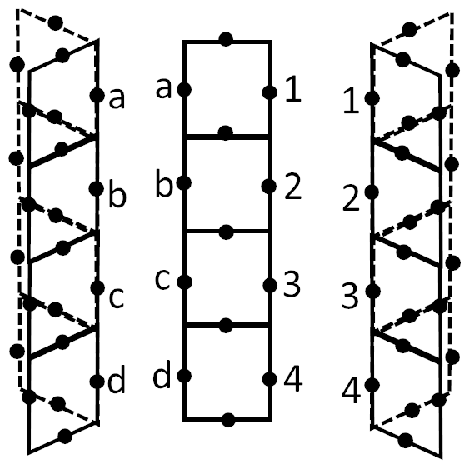}
         \caption{}\label{tc11}
          \end{subfigure}
  \hspace{10mm}
 \begin{subfigure}[h]{0.16\textwidth}
       \includegraphics[width=1.0\textwidth]{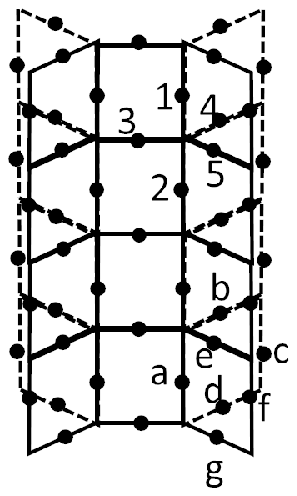}
         \caption{}\label{tc22}
          \end{subfigure}
  \hspace{10mm}
   \begin{subfigure}[h]{0.16\textwidth}
       \includegraphics[width=1.8\textwidth]{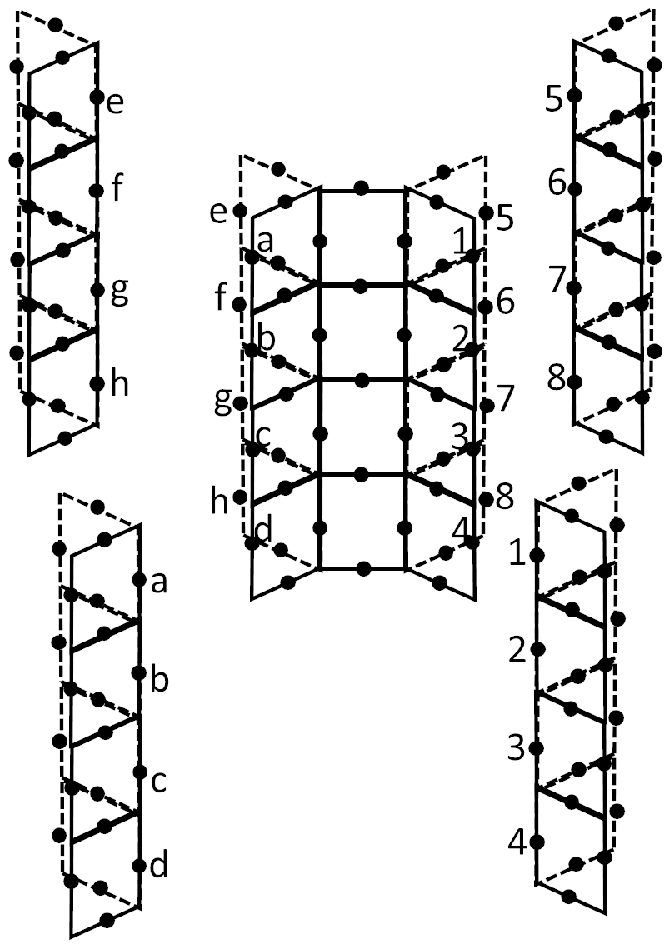}
         \caption{}\label{tc2210}
          \end{subfigure}

\begin{subfigure}[h]{0.13\textwidth}
    \includegraphics[width=1.0\textwidth]{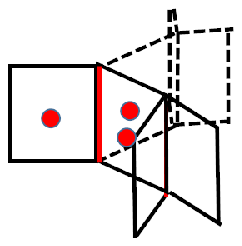}
         \caption{}\label{z2c}
             \end{subfigure}
             \begin{subfigure}[h]{0.13\textwidth}
    \includegraphics[width=1.0\textwidth]{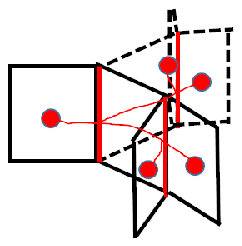}
         \caption{}\label{z2d}
             \end{subfigure}
              \begin{subfigure}[h]{0.10\textwidth}
    \includegraphics[width=1.0\textwidth]{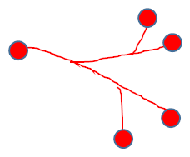}
         \caption{}\label{z2e}
             \end{subfigure}
 \end{center}
       
  \caption{ 
      (a)-(c) Construction of the lattice, which is referred to as the ``book-page" lattice in this paper. Qubits with the same numbers or letters are identified.
      (d) When applying a $X$ operator at a single vertical link (red bold line), $k$ m-anyons (red dots) are created. (e)(f)~Trajectories of the m-anyons and its top view. }
        \label{z2}
   \end{figure}
   Let us start with introducing our lattice model. The model is defined on a lattice constructed as a tensor product of 
  the Cayley tree with coordination number $k$ and a 1D chain, which is refereed to as the book-page lattice throughout this paper. 
   The lattice is constructed as follows: Starting with a ladder lattice, that we call spine, (the middle of Fig.~\ref{tc11}), we consider gluing two new ``branches" (more precisely, lattices consisting of tensor product of the branches of the Cayley tree and 1D lattice), which corresponds to the left- and right-most of Fig.~\ref{tc11}. We dub such branches books. 
   As an example, in the case of $k=3$, each book has two ``pages". After gluing books to the spine, we have the configuration shown in Fig.~\ref{tc22}, forming the book-page lattice up to the first generation\footnote{ For the sake of convenience, we slightly modify the notion of the generation of the book-page lattice in contrast to the one which is widely used in the Cayley tree.}. 
   This procedure can be iterated to construct the next generation; in the case of $k=3$, we attach four new books, each of which has two pages, to the end of the first generation, forming the second generation of the book-page lattice, as depicted in Fig.~\ref{tc2210}.
    The lattice model with 
    generic values of branches~($k>2$)\footnote{The case with $k=2$ corresponds to the 2D plane.}, and generation $M$ is similarly constructed.

Having defined the book-page lattice, 
we put the $\mathbb{Z}_2$ toric code~\cite{Kitaev2003} on the lattice. Introducing a qubit on each link (black dots in Fig.~\ref{z2}),
we define the Hamiltonian as \begin{equation}
    H=-J_A\sum_vA_v-J_B\sum_pB_p,\label{toric code}
\end{equation}
where $A_v$ (vertex term) is defined by the multiplication of Pauli $X$ operators that act on the links connected with a given vertex $v$, and $B_p$ (plaquette term) is a product of four Pauli~$Z$ operators on a plaquette. (Here, we define the Pauli operators, $X$ and $Z$, by the operators acting on a qubit, subject to relations $X^2=Z^2=I$, $XZ=-ZX$ with $I$ being identity operator.)
Since we will discuss a closed surface of the book-page lattice in the next section, we introduce only bulk terms of the Hamiltonian, not boundary terms. In the case of $k=3$, one of $A_v$ is shown by $A_v=X_1X_2X_3X_4X_5$ in Fig.~\ref{tc22}. Each plaquette operator consists of four $Z$ operators (for instance, $Z_aZ_bZ_cZ_{d}$ and $Z_aZ_eZ_fZ_{g}$ in Fig.~\ref{tc22}). Pictorially, we draw $A_v$ defined at the vertex which has incident edges with indices $1\sim5$ and three $B_p$'s which have support
on the edge indexed by 1 in Fig.~\ref{tc22} as
\begin{equation}
A_v = 
\begin{tikzpicture}[baseline={([yshift=-.5ex]current bounding box.center)},vertex/.style={anchor=base,
    circle,fill=black!25,minimum size=18pt,inner sep=2pt},scale = 0.8pt]
 
 \draw (0,1)--(0,-1);
 \draw (-1,0)--(0,0)--(0.67,0.33);
 \draw (0,0)--(0.67,-0.33);
 
 \filldraw[blue] (-0.5,0) circle (2pt);
 \filldraw[blue] (0,0.5) circle (2pt);
 \filldraw[blue] (0,-0.5) circle (2pt);
 \filldraw[blue] (0.33,0.166) circle (2pt);
 \filldraw[blue] (0.33,-0.166) circle (2pt);

\end{tikzpicture}, \qquad B_p = \begin{tikzpicture}[baseline={([yshift=-.5ex]current bounding box.center)},vertex/.style={anchor=base,
    circle,fill=black!25,minimum size=18pt,inner sep=2pt},scale = 0.8pt]
 
 \draw (0,0)--(1,0)--(1,1)--(0,1)--cycle;
 \draw (2,0)--(2.67,0.33)--(2.67,1.33)--(2,1)--cycle; ,
 \draw (3.5,1)--(4.17,0.67)--(4.17,-0.33)--(3.5,0)--cycle;
 
 \filldraw[red] (0.5,0) circle (2pt);
 \filldraw[red] (0,0.5) circle (2pt);
 \filldraw[red] (1,0.5) circle (2pt);
 \filldraw[red] (0.5,1) circle (2pt);
 
  \filldraw[red] (2.33,1.167) circle (2pt);
 \filldraw[red] (2,0.5) circle (2pt);
 \filldraw[red] (2.33,0.167) circle (2pt);
 \filldraw[red] (2.67,0.833) circle (2pt);
 
  \filldraw[red] (3.5,0.5) circle (2pt);
 \filldraw[red] (3.833,0.833) circle (2pt);
 \filldraw[red] (4.17,0.167) circle (2pt);
 \filldraw[red] (3.833,-0.167) circle (2pt);

\end{tikzpicture},
\end{equation}
where blue (red) dots represent $X$ ($Z$) Pauli operators.\par
It is easy to verify that terms defined in \eqref{toric code} commute with each other. Assuming the amplitude $J_A$ and $J_B$ is large, the ground state~$\ket{\psi}$ satisfies $A_v\ket{\psi}=\ket{\psi},\;B_p\ket{\psi}=\ket{\psi},\;\forall A_v,B_p$.
Following~\cite{Kitaev2003}, 
 the ground state is the simultaneous $+1$-eigenstate of all of the mutually commuting terms~$A_v, B_p$, and these mutually commuting terms are called \textit{stabilizers}.
 %
  \par

There are two types of anyons --- ``electronic" and ``magnetic" charges, which we abbreviate as $e$- and $m$-anyons. They are excitations that are charged under the vertex and plaquette terms, respectively (\textit{i.e.,} that violate the invariance of the action under the vertex and plaquette terms ). The $e$-anyon is the same as the one in the toric code in the 2D plane in a way that it can be created by a pair when acting a single $Z$ operator on the ground state and deformable in the bulk. The $m$-anyon, however, shows unusual behaviours compared with the case of toric code on the 2D plane. Acting an $X$ operator at one vertical link of the model, as shown in Fig.~\ref{z2c}, then $k$ $m$-anyons are created. This process is schematically described by $I\to \underbrace{m\otimes m\otimes \cdots \otimes m}_\textit{k}$ with $I$ and $m$'s represent the vacuum sector and $k$ $m$-anyonic excitations created by a $X$-operator acting on a single vertical link.
Furthermore, successive actions of the $X$ operators, the trajectory of the $m$-anyons exhibits the Cayley tree pattern as demonstrated in Fig.~\ref{z2d} and~\ref{z2e}. In other words, the wavefuction of the $m$-anyons is delocalized, spatially spreading as the generation increases. Due to this unusual behaviour of the $m$-anyon, one naturally wonders whether topological properties of this model are qualitatively different from the ones in the case of $\mathbb{Z}_2$ topologically ordered phase on the regular plane. 
In what follows, we confirm this intuition by investigating topological properties 
from two perspectives: the analyses of the GSD and the entanglement entropy. 
Due to unusual behavior of $m$-anyons and the hyperbolic geometry, one would not be able to see fractional statistics between $e$- and $m$-anyons immediately. In later section~(Sec.~\ref{33}), we give 
more thorough discussions on fractional statistics of these excitations. 
\par

\section{Ground States on a Closed surface }\label{sec tor}
When studying a topologically ordered phase, 
one would ask what is the GSD when we put it on a closed manifold. For instance, 
when a topologically ordered phase is placed on a torus, the GSD becomes non-trivial and
the number of distinct superselection sectors is equal to the GSD~\cite{Elitzur:1989nr}. 
We discuss the GSD on a closed book-page lattice and its relation with distinct anyonic excitations.
After having the ground states, we construct operators acting on these states to extract fractional statistics between quasiparticle excitations. We also discuss the stability of the GSD. 


\begin{figure}
 
\begin{center}
   \begin{subfigure}[h]{0.16\textwidth}
       \includegraphics[width=1.7\textwidth]{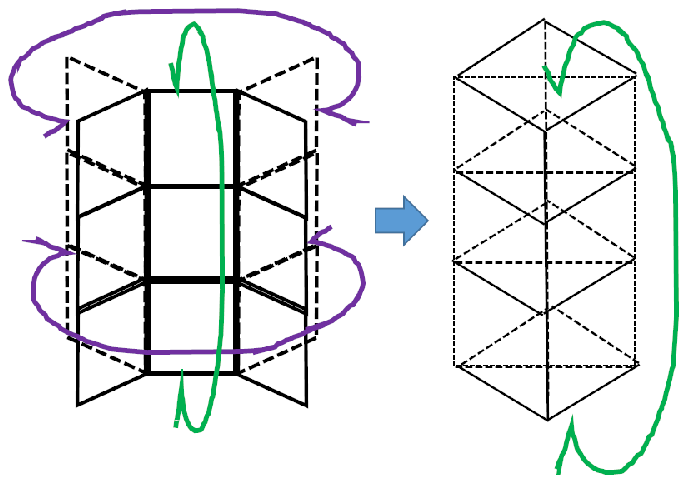}
         \caption{}\label{bpt}
          \end{subfigure}
  \hspace{20mm}
\begin{subfigure}[h]{0.10\textwidth}
    \includegraphics[width=2\textwidth]{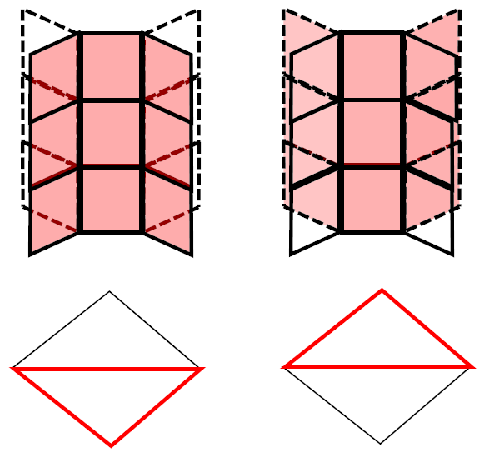}
         \caption{}\label{closed}
             \end{subfigure}
    \hspace{20mm} 
             \begin{subfigure}[h]{0.19\textwidth}
 \includegraphics[width=1.5\textwidth]{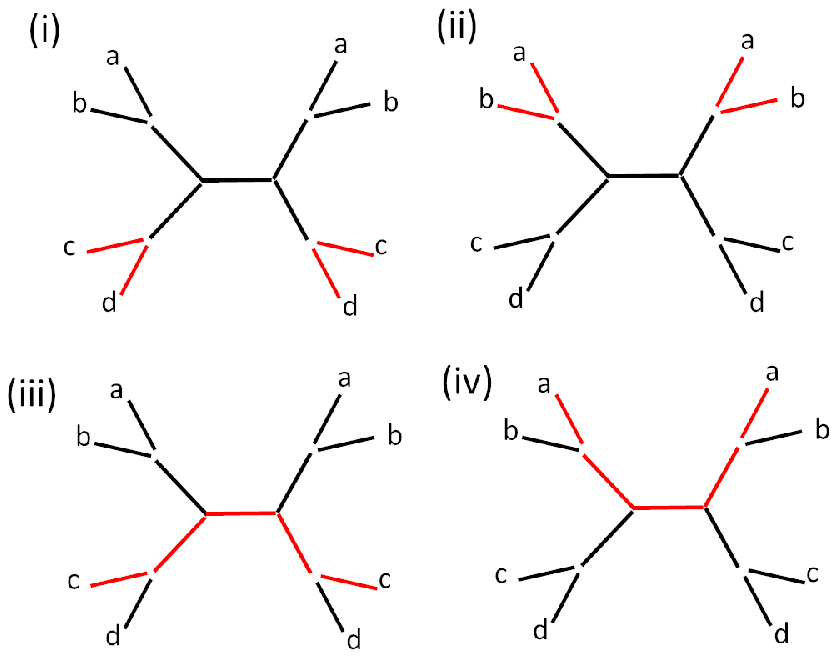}
         \caption{}\label{bpt3}
             \end{subfigure}
             \end{center}
             
             \begin{center}
\begin{subfigure}[h]{0.19\textwidth}
 \includegraphics[width=0.9\textwidth]{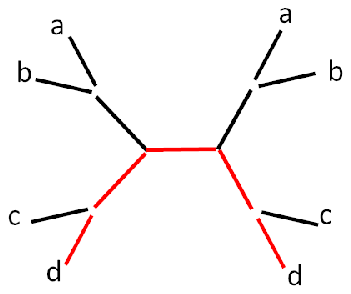}
         \caption{}\label{bpt8}
             \end{subfigure}
             \end{center}
      
  \caption{ (a) Construction of a closed surface of our model ($k=3$, $L=3$ and $M=1$). In the horizontal direction, the links at the end of the book-page lattice on the right are identified with the ones on the left (purple arrows). The periodic boundary condition is imposed in the vertical direction, indicated by a green arrow.  
      (b) Top: constraints on the stabilizers $B_p$ in the same geometry as (a) (the same boundary condition is imposed). The multiplication of plaquette operators in region marked by red gives identity. There are two such constraints in this case. Bottom: the top view of the closed surface. The red line corresponds to the top view of trajectory of plaquette terms depicted above. 
      (c) The top view of closed surface in the case of $k=3$ and $M=2$. The red lines represent the top view of trajectories of plaquette terms, analogously to the ones in the bottom of (b). The same boundary condition is imposed as (a) so that vertical links with the same alphabet are identified. (d) The top view of one close surface that can be generated by (i) and (iii) in (c).
      }
        \label{z33}
   \end{figure}
   
\subsection{Counting GSD}
We start with the $\mathbb{Z}_2$ toric code on the book-page lattice with finite size up the $M$ generation and height being $L$ (\textit{i.e.,} the lattice contains $L$ vertical links). Since the geometry is symmetric with respect to the spine, we can close the lattice in the horizontal direction by identifying the end points on the right and those on the left. Further, we also identify the endpoints of at the top boundary and those at the bottom boundary. These procedures yield a closed surface which is referred to as the ``book-page torus". 
We portray one example of the book-page torus in Fig.~\ref{bpt}.\par
The GSD is equal to $2^{N_q-N_s}$ with $N_q$ and $N_s$ being the number of qubits and independent stabilizers, respectively. 
The number of qubits in this closed surface is given by
\begin{equation}
    {N_q}=3L+4L[(k-1)+\cdots +(k-1)^{M-1}]+3L(k-1)^M .\label{nq}
\end{equation}
The number of independent stabilizers is equal to the number of individual stabilizers defined in \eqref{toric code}
subtracted by the number of constraints that the multiplications of stabilizers give. It is straightforward to show that the number of individual stabilizes, $A_v$ and $B_p$ in \eqref{toric code} equals the one given in~\eqref{nq}, as it is in the same way of the toric code on a torus.\par
As for the constraints, the multiplication of all the vertex terms $A_v$ gives the identity, yielding one constraint. The multiplication of the plaquette terms $B_p$
which form a closed loop along the horizontal direction also gives the identity. Examples are exhibited in Fig.~\ref{closed}. Due to the nature of Cayley tree which has $k$ branches at each node, there are a number of distinct loops in the book-page torus, leading to a number of constraints involving the plaquette terms. 
Thus, the number of such constraints involving plaquette terms depends on the number of distinct loops in the horizontal direction. 
Let us 
look at simple examples.
In the case of $M=1$, there are $k-1$ such loops (Fig.~\ref{closed}). 
In the case of $M=2$, there are $(k-1)(k-2)$ distinct loops involving 
links which belongs to only the second generation. [In the case of $k=3$, there are two such loops corresponding to (i) and (ii) in Fig.~\ref{bpt3}.] Also, there are $k-1$ distinct loops that involve links at the all of the generations. [In the case of $k=3$, there are two such loops corresponding to (iii) and (iv) as indicated in in Fig.~\ref{bpt3}.] It is important to note that all loops in the horizontal direction with $M=2$ can be generated by these loops. [As an instance with $k=3$, a loop portrayed in Fig.~\ref{bpt8} can be generated by those of (i) and (iii) in Fig.~\ref{bpt8}.] 
In total, there are $(k-1)(k-2)+(k-1)=(k-1)^2$ distinct loops in the horizontal direction, hence, the number of constraint on multiplication of $B_p$'s is also given by $(k-1)^2$. This line of thoughts can be generalized to any number of $M$.
The number of distinct loops in the horizontal direction, equivalently, the number of constraints on the plaquette terms, reads
\begin{equation*}
    (k-1)^{M-1}(k-2)+(k-1)^{M-2}(k-2)+\cdots+(k-1)=(k-1)^M.
\end{equation*}
The number of the independent stabilizers then gives
\begin{equation*}
    N_s=N_q -1 - (k-1)^M,
\end{equation*}
hence\footnote{When $k=2$ the closed surface become the regular torus, giving the $GSD=2^2$, which is consistent with the well-known result of the $\mathbb{Z}_2$ toric code on the torus.}~\footnote{The sub-extensive GSD can be seen in fracton topological phases, which are topological phases exhibiting unusual GSD dependence on the UV lattice spacing
~\cite{chamon,Haah2011,Vijay}. Here, the result~\eqref{gsd} also shows the GSD dependence on the system size of the lattice, yet
the sub-extensive GSD of our model has the qualitatively different origin from the fracton: while in the fracton topological phases, mobility constraint on excitation leads to the sub-extensive GSD,  
in our case, 
the geometric structure of the Cayley tree 
gives rise to the sub-extensive GSD.}
\begin{equation}
    GSD=2^{N_q-N_s}=2^{1+(k-1)^{M}}.\label{gsd}
\end{equation}

\par
 \begin{figure}
  \begin{center}
   \includegraphics[width=0.95\textwidth]{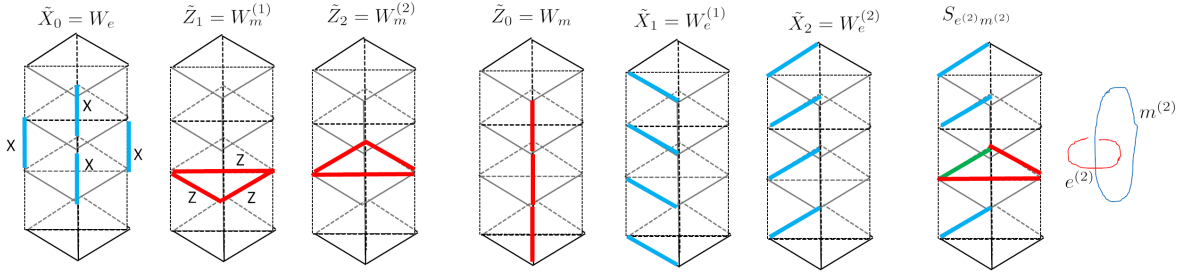}
        \end{center}
\caption{(The first to sixth) Logical operators, which are equivalent to flux operators that measure magnetic/electric charges threading in the vertical and horizontal direction with $k=3$ and $M=1$. (seventh and eighth) Mutual braiding between $e^{(2)}$ and $m^{(2)}$, corresponding to $(3,3)$ component of the $S$-matrix~\eqref{smatrix}. }\label{ms}
\end{figure}
%

\subsection{Explicit form of the ground states} 
Following the previous discussions on the GSD, in this subsection, we
give an alternative interpretation on the result~\eqref{gsd} by explicitly writing down the ground states distinguished by non-contractible closed loops of anyons. For the sake of the simplicity of terminology, we refer such closed loops of anyons to as \textit{logical operators}, borrowing a jargon in the context of quantum information~\cite{Kitaev2003}.  
To 
start, 
let $\mathcal{H}$ be the Hilbert space defined by $\mathcal{H}=\mathcal{H}_{1/2}^{\otimes N_q}$, where $\mathcal{H}_{1/2}=\text{span}\{\ket{0},\ket{1}\}$, \textit{i.e.,} tensor product of spin-$1/2$ states ($\ket{0}$/$\ket{1}$ corresponds to spin-up/-down state in the spin-$z$ basis) on links of the lattice. Define the following state
\begin{equation}
    \ket{\Psi}=\frac{1}{\sqrt{2^{N_{p}}}}\prod_{1\leq t\leq N_p} (1+B_{p_t})\ket{\to}^{\otimes N_q}\label{gsd4}
\end{equation}
 with $\ket{\to}$ being the diagonal basis of the $X$ operator: $\ket{\to}=\frac{1}{\sqrt{2}}(\ket{0}+\ket{1}). $\footnote{We have the prefactor in \eqref{gsd4} to normalized the state to ensure the norm is unit.} $N_p$ is the number of independent plaquettes in the model and $B_{{p}{t}}$ denotes the plaquette term at plaquette~$p_t$ introduced in Hamiltonian~\eqref{toric code}.
One can verify that \eqref{gsd4} satisfies $A_v\ket{\Psi}=\ket{\Psi},\;B_p\ket{\Psi}=\ket{\Psi},\;\forall A_v,B_p$.
The $2^{F_{M,k}}$ ground states are characterized by $2^{F_{M,k}}$ pairs of logical operators   $(\tilde{Z}_i,\tilde{X}_i)$ $(0\leq i\leq F_{M,k}-1)$ which satisfy 
\begin{equation}
\{\tilde{X}_{i},\tilde{Z_i}\}=0,\;[\tilde{X}_{i},\tilde{Z_j}]=0\;(i\neq j).\label{stab}
\end{equation}
\par
 
Let us first look at the form of the logical operators $\tilde{Z_i}$, which correspond to the non-contractible loops of $e$-anyon. 
Logical operators are counted as equivalent if they differ by stabilizers. In this view, 
there is only one logical operator $\tilde{Z}_0$ that goes along the vertical direction whereas there are $F_{M,k}-1=(k-1)^M$ distinct logical operators $\tilde{Z}_{i}\;(1\leq i\leq F_{M,k}-1)$ in the horizontal direction, which is in line with the fact that there are $(k-1)^M$ distinct loops in the horizontal direction. 
This is contrasted with the case of $\mathbb{Z}_2$ toric code on the regular torus, where there is only one logical operator that wraps around the torus in each direction. Examples are shown in Fig.~\ref{ms}. \par
We also consider the form of logical operators $\tilde{X_i}$, which is in the dual description of what we have discussed in the previous paragraph.  These logical operators correspond to the non-contractible loops of $m$-anyons. There is only one logical operator $\tilde{X}_0$ that goes along the horizontal direction
involving the all the vertical links along the trajectory, forming a membrane shape (see the first geometry in Fig.~\ref{ms}). Also, 
there are $F_{M,k}-1=(k-1)^M$ distinct logical operators $\tilde{Z}_{i}\;(1\leq i\leq F_{M,k}-1)$ in the vertical direction. 
One can show these logical operators indeed satisfy \eqref{stab} and using these pairs of logical operators and the stabilized state~\eqref{gsd4},
the $2^{F_{M,k}}$ ground states are defined by
\begin{equation}
    \ket{\varphi_{a_0,(a_1,\cdots,a_{F_{M,k}-1})}}=\prod_{0\leq i\leq F_{M,k}-1}\tilde{Z}_i^{a_i}\ket{\Psi}\label{ket}
\end{equation}
with $a_i=0,1\;(\text{mod}\;2)$. 
\subsection{$S$ and $T$ matrices}\label{33}
After defining the ground states~\eqref{ket}, one can introduce $S$ and $T$ matrices that map between ground states in the book-page torus. Similar to the modular $S$ and $T$ matrix of the toric code on the regular torus, these matrices convey anyonic properties: mutual braiding statistics and self-statistics, associated with topological spins~\cite{Kitaev2006}.
 In this subsection, we discuss the case of $k=3$ and $M=1$.
 The generalization to the other values of $k$ and $M$ is straightforward and is briefly mentioned in the end of this subsection.


To proceed, we construct ground states which carry electronic or magnetic charges threading into either horizontal or vertical direction of the book-page torus. These states are called minimal entangled states, which can be associated with superselection sectors of a topologically ordered phase~\cite{grover2012}. 
Let us first construct ground states carrying electric or magnetic charges threading into the vertical direction. 
Define flux operators which are non-contractible loops along the horizontal direction, $W_e$, $W_m^{(1)}$, $W_m^{(2)}$, the first (last two) of which measures the electric (magnetic) flux threading in the vertical direction defined by multiplication of $X$ operators ($Z$ operators), see first three geometries in Fig.~\ref{ms}. It is straightforward to show that 
\begin{eqnarray}
     W_e\ket{\varphi_{a_0,(a_1,a_2)}}&=&(-1)^{a_0}\ket{\varphi_{a_0,(a_1,a_2)}}\\W_m
   ^{(1)}\ket{\varphi_{a_0,(a_1,a_2)}}&=&\ket{\varphi_{a_0,(a_1+1,a_2)}}\\
   W_m
   ^{(2)}\ket{\varphi_{a_0,(a_1,a_2)}}&=&\ket{\varphi_{a_0,(a_1,a_2+1)}}.
\end{eqnarray}
One can introduce the eigenstates of these flux operators:
\begin{eqnarray}
    &&\ket{I}_v=\frac{1}{2}\sum_{a_1,a_2}\ket{\varphi_{0,(a_1,a_2)}},\;\;\; \ket{e}_v= \frac{1}{2}\sum_{a_1,a_2}\ket{\varphi_{1,(a_1,a_2)}} ,\;\;\;
     \ket{m^{(i)}}_v=\frac{1}{2}\sum_{a_1,a_2}(-1)^{a_i}\ket{\varphi_{0,(a_1,a_2)}}\nonumber\\
     &&\ket{m^{(1)}m^{(2)}}_v=\frac{1}{2}\sum_{a_1,a_2}(-1)^{a_1+a_2}\ket{\varphi_{0,(a_1,a_2)}},\;\;\;
     \ket{em^{(i)}}_v=\frac{1}{2}\sum_{a_1,a_2}(-1)^{a_i}\ket{\varphi_{1,(a_1,a_2)}}\nonumber\\
     &&\ket{em^{(1)}m^{(2)}}_v=\frac{1}{2}\sum_{a_1,a_2}(-1)^{a_1+a_2}\ket{\varphi_{1,(a_1,a_2)}},\;\;(i=1,2).\label{eight1}
    \end{eqnarray}
These eight states exhaust all the ground states which carry charges in the vertical direction measured by the flux operators. 
The meaning of the labels of the states in \eqref{eight1} is clear from the context. For instance, the last state is the generalized dyon, carrying one electric charge and two magnetic charges, measured by the three operators, $W_e$, $W_m^{(1)}$, $W_m^{(2)}$.
\par
Similarly, one can construct ground states carrying electronic and magnetic charges along the horizontal direction. Introducing flux operators $ W_m$, $W_e
   ^{(1)}$, and $W_e
   ^{(2)}$ wrapping around the vertical direction as shown in the last three geometries in Fig.~\ref{ms}, their actions on the ground states are given by
\begin{eqnarray}
    W_m\ket{\varphi_{a_0,(a_1,a_2)}}&=&\ket{\varphi_{a_0+1,(a_1,a_2)}}\\W_e
   ^{(1)}\ket{\varphi_{a_0,(a_1,a_2)}}&=&(-1)^{a_1}\ket{\varphi_{a_0,(a_1,a_2)}}\\
   W_e
   ^{(2)}\ket{\varphi_{a_0,(a_1,a_2)}}&=&(-1)^{a_2}\ket{\varphi_{a_0,(a_1,a_2)}}.
\end{eqnarray}
Following eight states are eigenstates of these operators:
\begin{eqnarray}
    &&\ket{I}_h=\frac{1}{\sqrt{2}}[\ket{\varphi_{0,(0,0)}}+\ket{\varphi_{1,(0,0)}}],\;\;\;
     \ket{m}_h=\frac{1}{\sqrt{2}}[\ket{\varphi_{0,(0,0)}}-\ket{\varphi_{1,(0,0)}}]\nonumber\\
    && \ket{e^{(1)}}_h=\frac{1}{\sqrt{2}}[\ket{\varphi_{0,(1,0)}}+\ket{\varphi_{1,(1,0)}}],\;\;\;
      \ket{e^{(2)}}_h=\frac{1}{\sqrt{2}}[\ket{\varphi_{0,(0,1)}}+\ket{\varphi_{1,(0,1)}}]\nonumber\\
      &&  \ket{e^{(1)}e^{(2}}_h=\frac{1}{\sqrt{2}}[\ket{\varphi_{0,(1,1)}}+\ket{\varphi_{1,(1,1)}}],\;\;\;
     \ket{e^{(1)}m}_h=\frac{1}{\sqrt{2}}[\ket{\varphi_{0,(1,0)}}-\ket{\varphi_{1,(1,0)}}]\nonumber\\
     &&\ket{e^{(2)}m}_h=\frac{1}{\sqrt{2}}[\ket{\varphi_{0,(0,1)}}-\ket{\varphi_{1,(0,1)}}],\;\;\;
      \ket{e^{(1)}e^{(2)}m}_h=\frac{1}{\sqrt{2}}[\ket{\varphi_{0,(1,1)}}-\ket{\varphi_{1,(1,1)}}].\label{eight2}
    \end{eqnarray}
    The states \eqref{eight1} and \eqref{eight2} correspond to the superselection sectors in the vertical and horizontal direction. Having introduced these states, one can construct the $S$ matrix that maps superselection sectors~\eqref{eight1} to the ones~\eqref{eight2}:
\begin{equation}
    S_{[h],[v]}=\frac{1}{2\sqrt{2}}\begin{pmatrix}
1 & 1 & 1&1&1&1&1&1\\
1 & -1 & 1&-1&1&-1&1&-1\\
1 & 1 & -1&-1&1&1&-1&-1\\
1 & -1 & -1&1&1&-1&-1&1\\
1 & 1 & 1&1&-1&-1&-1&-1\\
1 & -1 & 1&-1&-1&1&-1&1\\
1 & 1 & -1&-1&-1&-1&1&1\\
1 & -1 & -1&1&-1&1&1&-1
\end{pmatrix}.\label{smatrix}
\end{equation}
Here, the column (row) $[h]$ ($[v]$) denotes the superselection sectors in the horizontal (vertical) direction described by $[h]=(I,e^{(1)},e^{(2)},e^{(1)}e^{(2)},m,e^{(1)}m,e^{(2)}m,e^{(1)}e^{(2)}m)$\par\noindent ($[v]=(I,m^{(1)},m^{(2)},m^{(1)}m^{(2)},e,em^{(1)},em^{(2)},em^{(1)}m^{(2)})$).
Similar to the modular $S$-matrix of the $\mathbb{Z}_2$ toric code on the regular torus, the sign of each component of the matrix~\eqref{smatrix} characterizes the mutual braiding statistics between superselection sectors in the different direction. As an example, the sign of the $(3,3)$ component of this matrix is negative, which is consistent with the fact that $e^{(2)}$ in the vertical direction has $\pi$ mutual statistics when braided around magnetic charge $m^{(2)}$ in the horizontal direction. Visualization of such mutual braiding is demonstrated in the last two geometries in Fig.~\ref{ms}.
\par
One can also introduce the $T$ matrix, another important matrix characterizing topological spin, corresponding to the self-statistics of an anyon. The operation of the $T$ matrix is equivalent to the Dehn twist, accomplished by moving logical operator $\tilde{Z}_0$ around closed loop in the horizontal direction. Since there are $(k-1)^M$ distinct loops in the horizontal direction, there are $(k-1)^M$ $T$ matrices, denoted by $T^{(i)}\;[1\leq i\leq (k-1)^M]$. In the case of $k=3$ and $M=1$, there are two $T$-matrices, $T^{(1)}$ and $T^{(2)}$, we schematically exhibit the operation of the $T^{(1)}$ in Fig,~\ref{dehn}. These operations acts on the ground state~\eqref{ket} as
\begin{equation}
    T^{(1)}\ket{\varphi_{a_0,(a_1,a_2)}}=\ket{\varphi_{a_0,(a_0+a_1,a_2)}},\;\; T^{(2)}\ket{\varphi_{a_0,(a_1,a_2)}}=\ket{\varphi_{a_0,(a_1,a_0+a_2)}}.
\end{equation}
Therefore, in the basis of the superselection sector in the vertical direction,\par

\noindent $[v]=(I,m^{(1)},m^{(2)},m^{(1)}m^{(2)},e,em^{(1)},em^{(2)},em^{(1)}m^{(2)})$, the $T$ matrices are given by
\begin{eqnarray}
    T^{(1)}&=&\text{diag}(1,1,1,1,1,-1,1,-1)\\
     T^{(2)}&=&\text{diag}(1,1,1,1,1,1,-1,-1).
\end{eqnarray}

\par
The generalization to other values of $k$ and $M$ is straightforward. In the horizontal direction, there are $F_{M,k}$ different charges $m,e^{(i)}\;(1\leq i\leq F_{M,k})$ with which $2^{F_{M,k}}$ superselection sectors are generated. Likewise, in the vertical direction, there are $F_{M,k}$ different charges  $e,m^{(i)}\;(1\leq i\leq F_{M,k})$, allowing us to have $2^{F_{M,k}}$ superselection sectors. Introducing two vectors $\bm{v}=(v_1\cdots,v_{F_{M,k}})$ and $\bm{h}=(h_1\cdots,h_{F_{M,k}})$, where each entry takes either $0$ or $1$,
the superselection sectors in the horizontal and vertical direction are defined by 
\begin{eqnarray}
 \ket{\bm{h}}&=&\ket{m^{h_1}(e^{(1)})^{h_2}\cdots (e^{(F_{M,k})})^{h_{F_{M,k}}}}\nonumber\\
 \ket{\bm{v}}&=&\ket{e^{v_1}(m^{(1)})^{v_2}\cdots (m^{(F_{M,k})})^{v_{F_{M,k}}}},
\end{eqnarray}
the $S$ matrix is given by
\begin{equation}
    S_{[\bm{v}],[\bm{h}]}=\frac{1}{\sqrt{2^{F_{M,k}}}}(-1)^{\bm{v}\cdot\bm{h}}.
\end{equation}
There are $(k-1)^M$ $T$ matrices, $T^{(j)}\;(1\leq j\leq F_{M,k}-1)$ whose action on the superselection sectors in the vertical direction reads
\begin{equation}
    T^{(j)}\ket{\bm{v}}=(-1)^{v_1\cdot v_{j+1}}\ket{\bm{v}}.
\end{equation}

\subsection{Stability}\label{sec stability}
As \eqref{gsd} shows, the GSD grows exponentially with the generation $M$
of the book-page lattice. In this subsection, we discuss stability of these ground states under local perturbations. As we mentioned previously, the $2^{F_{M,k}}=2^{1+(k-1)^M}$ ground states are characterized by $F_{M,k}$ pairs of logical operators, $\tilde{X}_{i}/\tilde{Z_i}$ $(0\leq i\leq F_{M,k}-1)$ 
with which one can associate non-contractible loops of $m/e$ anyons, 
satisfying $\{\tilde{X}_{i},\tilde{Z_i}\}=0$, $[\tilde{X}_{i},\tilde{Z_j}]=0\;(i\neq j)$. As we have depicted logical operators in Fig.~\ref{ms},  
$\tilde{X}_0$ goes along the horizontal direction, forming a membrane shape whereas $\tilde{Z}_0$ is non-contractible loops of $e$-anyon going along the vertical direction. Regarding other pairs, $(\tilde{X}_{i}/\tilde{Z}_i)\;(i\neq 0)$, $\tilde{X}_i$ forms a loops of $e$-anyon in the horizontal direction and $\tilde{Z}_i$ is a non-contractible loop in the vertical direction. \par
For simplicity, in this subsection, we assume 
that the length of the book-page torus in the vertical direction is sufficiently large, i.e.,  
$L\gg 1$ so that logical operators $\tilde{Z}_0$
and $\tilde{X}_{i}\;(i\neq 0)$ are stable against a local perturbation $\lambda X_a$ or $\lambda Z_a$ with $\frac{|\lambda|}{|J|}\ll 1$. [Here, $J$ represents the amplitude of the vertex and plaquette terms in \eqref{toric code}, which is therefore of order of energy gap of the system.] 
Hence, we concentrate on the stability of the logical operators $\tilde{X}_0$
and $\tilde{Z}_{i}\;(i\neq 0)$, i.e., logical operators going along the horizontal direction with small number of genera ion. \par
\begin{figure}
    \begin{center}
     \begin{subfigure}[h]{0.5\textwidth}
    \includegraphics[width=1.0\textwidth]{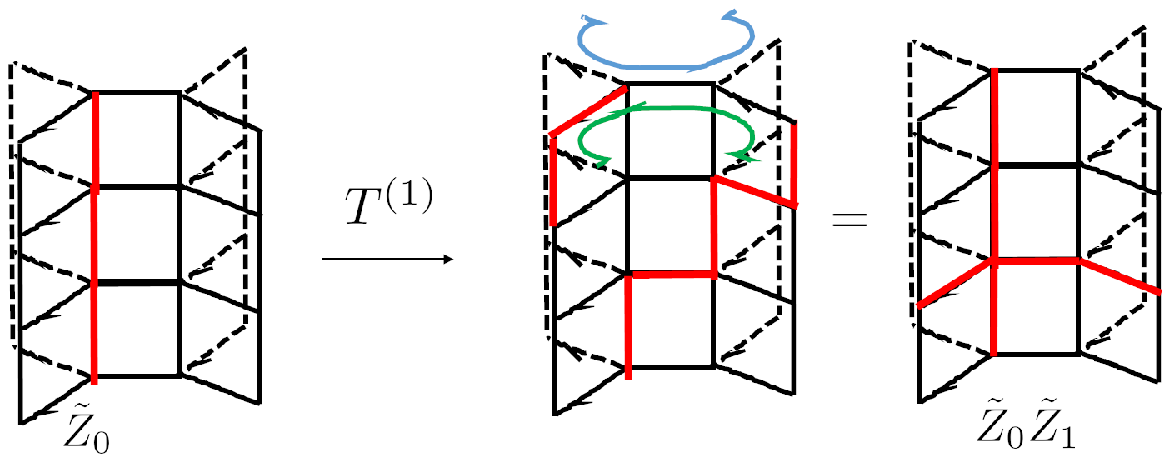}
         \caption{}\label{dehn}
             \end{subfigure}

             \hspace{10mm}
             
    \begin{subfigure}[h]{0.35\textwidth}
  \includegraphics[width=1.5\textwidth]{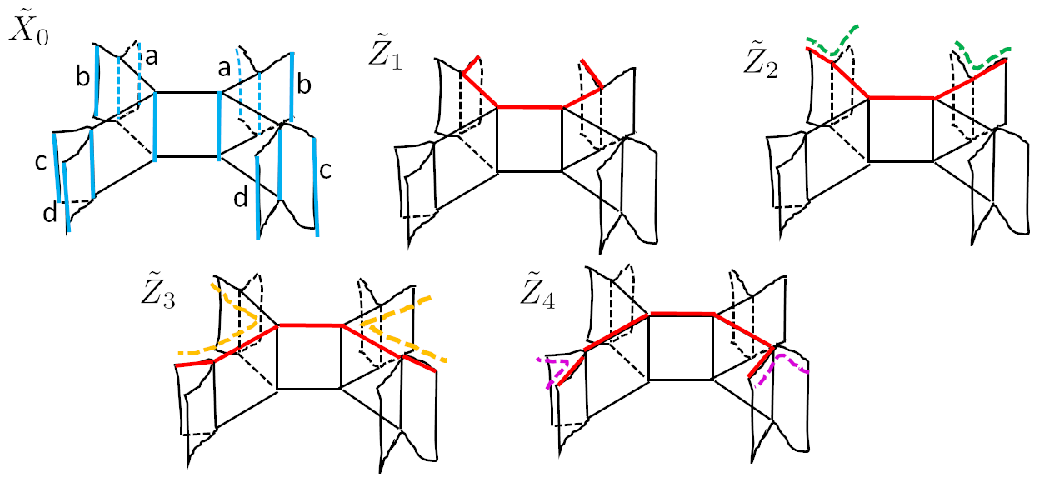}
     \caption{}\label{stability}
           \end{subfigure}
           \end{center}
               \caption{(a) Dehn twist in the book-page torus with $k=3$ and $M=1$, corresponding to $T^{(1)}$. The boundary condition is imposed in the same manner as Fig.~\ref{bpt}.
               The logical operator $\tilde{Z}_0$ moves around a loop (green arrow). Other Dehn twist, corresponding to $T^{(2)}$ is similarly implemented where $\tilde{Z}_0$ moves around the loop marked by blue arrow.
               (b) One logical operator $\tilde{X}_0$ and four logical operators $\tilde{Z}_i$ going along the horizontal direction in the book-page torus with $k=3$ and $M=2$. The periodic boundary condition is imposed in such a way that links with the same alphabet are identified.}
 \end{figure}

In the case of $k=3$ and $M=2$, the GSD is given by $2^5$ from Eq.~(\ref{gsd}). Accordingly, there are one logical operator $\tilde{X}_0$ and four logical operators $\tilde{Z}_1$, $\tilde{Z}_2$, $\tilde{Z}_3$, and $\tilde{Z}_4$ going along the horizontal direction. We show these operators in Fig.~\ref{stability}.
The most stable logical operator among them is $\tilde{X}_0$,  whose trajectory contains all of the vertical links. As for $\tilde{Z}_i\;(1\leq i\leq 4)$, it is important to note that $\tilde{Z}_1$ and $\tilde{Z}_2$ ($\tilde{Z}_3$ and $\tilde{Z}_4$) differ by the small loop indicated by green (purple) dashed line in Fig.~\ref{stability} whose perimeter is four. Introducing a local spin-flip perturbation, which has the form $H_{\lambda}=\lambda\sum _aX_a$ with $\frac{|\lambda|}{|J|}\ll 1$, the degeneracy between ground states characterized by $\tilde{Z}_1$ and $\tilde{Z}_2$ as well as $\tilde{Z}_3$ and $\tilde{Z}_4$ is lifted by the fourth order perturbation~$\sim\bigl(\frac{\lambda}{J}\bigr)^4$.
It follows that the GSD is reduced to $2^3$. 
Eighth order perturbation~$\sim\bigl(\frac{\lambda}{J}\bigr)^8$ would lift the degeneracy characterized by $\tilde{Z}_2$ and $\tilde{Z}_3$, further
reducing the GSD to $2^2$. 
This is in line with the fact that ground states characterized by $\tilde{Z}_2$ and $\tilde{Z}_3$ differ by the bigger loop [orange dashed line in Fig.~\ref{stability}] are lifted by the eighth order perturbation. The remaining $2^2$ GSD coincides with the one of the $\mathbb{Z}_2$ toric code on the regular torus. 
Based on the argument given above, we factorize the $2^5$ GSD as
\begin{equation}
    2^5=2^2\cdot2^1\cdot2^2.\label{25}
\end{equation}
The first factor of \eqref{25} corresponds to the most stable ground states, which is the same as the toric code on the regular torus, whereas the second (third) factor of \eqref{25} ensures the increase of the GSD due to the presence of the large (small) loop marked by orange (green and purple) dashed line which is stable up to the higher order of perturbation. \par
As for the general cases, the $2^{F_{M,k}}=2^{1+(k-1)^M}$ GSD can be factorized into $M$ components:
\begin{equation}
     2^{1+(k-1)^M}=2^2\cdot2^{(k-2)}\cdot2^{(k-1)(k-2)}\cdots 2^{(k-1)^{M-1}(k-2)}.\label{gsd de}
\end{equation}
The first factor corresponds to the most stable ground states, and $r$-th ($2\leq r \leq M$) component of~\eqref{gsd de} corresponds to the GSD which is stable against perturbation up to the $4(M+1-r)$-th order $\sim\bigl(\frac{\lambda}{J}\bigr)^{4(M+1-r)}$. As $r$ increases, the loops are closer to the outer edge of the system, and less stable due to the shorter perimeter.

\section{Entanglement entropy}\label{nonlocal enta}
We turn to the second topological quantity, the topological entanglement entropy developed in~\cite{kim2014,PhysRevB.97.144106}. 
Topological entanglement entropy was initially discussed in~\cite{levinwen2006,preskillkitaev2006}. It captures universal properties of a topologically ordered phase. Recently, it was found that by making use of quantum information tools, the topological entanglement entropy conveys a clearer physical intuition -- the number of distinct anyonic excitations~\cite{kim2014,PhysRevB.97.144106}. In this section, we resort to the argument given in~\cite{kim2014,PhysRevB.97.144106} to study the interplay between anyonic excitations and geometric properties of the Cayley trees. We further present consistency checks of our result by introducing gapped boundary.
\subsection{Non-local entanglement entropy}
Before doing so, 
we briefly review the topological entanglement entropy developed in~\cite{kim2014,PhysRevB.97.144106}. 
Entanglement entropy in various topologically ordered phases has been intensively studied for years. For more detailed introductory discussions of this subject, readers may consult literature~\cite{Hamma2005,levinwen2006,preskillkitaev2006}. 
The basic idea is that conditional mutual information 
has an intimate relation with topological entanglement entropy if one sets properly the geometry of disjoint regions.\par 
Let us start by introducing the conditional mutual information. 
For a state $\rho$ in disjoint regions $A,B,C$, 
the conditional mutual information $I(A:C|B)$ is given by \begin{equation}
    I(A:C|B)|_{\rho}=  S(\rho_{AB})+S(\rho_{BC})- S(\rho_B)-S(\rho_{ABC}),\label{gs}
\end{equation}
where $S(\rho_{*})$ denotes the von-Neumann entanglement entropy in a subsystem.
Physically, this quantity captures the change of the correlation between $A$ and $B$ with and without $C$.\footnote{This can be seen by recalling $ I(A:C|B)|_{\rho}=I(A:BC)|_{\rho}-I(A:B)|_{\rho}$, where $I(A:BC)|_{\rho}$ is mutual information defined by $I(A:BC)|_{\rho}=S(\rho_{A})+S(\rho_{BC})- S(\rho_{ABC})$. } For 
another state $\sigma$, with 
the properties $\rho_{AB}=\sigma_{AB}$ and $\rho_{BC}=\sigma_{BC}$, it follows that 
\begin{equation}
     I(A:C|B)|_{\rho}= I(A:C|B)|_{\sigma}+S(\sigma_{ABC})-S(\rho_{ABC})\geq S(\sigma_{ABC})-S(\rho_{ABC}).\label{mutual}
\end{equation}
The last inequality holds due to the positivity of the conditional mutual information (which comes from the strong-subadditivity of the entanglement entropy) and the lower bound is saturated when $I(A:C|B)|_{\sigma}=0$
.\par
While the relation~\eqref{mutual} holds in any quantum state, it has a particularly suggestive geometric interpretation in the context of topologically ordered phases.  
To see this, 
from now on
we focus on a quantum state in a topologically ordered phase and specifically set geometry of $A$, $B$, and $C$ in this phase.
As an example, for a given 2D Abelian non-chiral topologically ordered phases (more precisely Abelian topologically ordered phase described by a local commuting Hamiltonian),
such as the toric code on a plane, we introduce the four-partite system $ABCD$ as portrayed in 
Fig.~\ref{ee}, so that $D$ is complement of $ABC$. 
Setting the state $\rho$ as the ground state of this system, 
the conditional mutual information~\eqref{gs} is now written as 
\begin{equation}
     I(A:C|B)|_{\rho}=  S(\rho_{BC})+S(\rho_{CD})- S(\rho_B)-S(\rho_{D})\equiv \tilde{S}\label{stilde}
\end{equation}
Here we have assumed that $\rho_{ABCD}$ is a pure state.
By appropriate unitary operations, one can create an anyon whose trajectory runs cross $C$ from $D$, with its end points being located within $D$, giving an excited state $\sigma$.
Suppose 
we have sets of distinct excited states $\sigma_I$ ($I=1,\cdots,N$) such that $\sigma_{I AB}=\rho_{AB}$ and $\sigma_{I BC}=\rho_{BC}$. Here, being distinct means $\sigma_I\cdot\sigma_J=0\;(I\neq J)$. Introducing a mixed state $\sigma^\prime=\sum_Ip_I\sigma_I\;(\sum_Ip_I=1)$, we substitute this excited state into \eqref{mutual} ($\sigma\to\sigma^\prime$) and obtain
\begin{equation}
    \tilde{S}\geq S(\sigma^{\prime}_{ABC})-S(\rho_{ABC})=-\sum_Ip_I\log p_I.
\end{equation}
It can be shown that the lower bound of \eqref{mutual} is saturated when $p_I=1/N$, in which $A$ and $C$ are conditionally independent: $ I(A:C|B)|_{\sigma^\prime}=0$ and the lower bound of the conditional mutual information of the ground state is given by $\tilde{S}=\log N$~\cite{PhysRevB.97.144106}. 
Furthermore, $\tilde{S}$ is a universal number. In fact, 
an explicit calculation shows that $\tilde{S}$ is 
topological in the sense of being independent of the area terms~\cite{levinwen2006,preskillkitaev2006}.  
%
%
Following the terminology in \cite{PhysRevB.97.144106}, we term $\tilde{S}$ the non-local entanglement entropy. 
\par
In the case of the $\mathbb{Z}_2$ toric code on the 2D plane, there are four types of distinct anyons, represented by $\eta_{i,j}=(e)^i(m)^j\;(i,j=0,1)$, where $e/m$ denotes the electronic/magnetic anyon. In other words, there are two generators of  $\mathbb{Z}_2$ excitations, $e$- and $m$- anyons, with which any excitation can be created, thus there are $N=2^2=4$ distinct excitations. The non-local entanglement entropy is saturated and gives
$\tilde{S}=\log 2^2=2\log 2$. 
\par
One can generalize this argument to an arbitrary non-chiral $\mathbb{Z}_2$ topologically ordered phase. In the four-partite system $ABCD$, suppose there are 
$n$ generators of anyons carrying $\mathbb{Z}_2$ charge and running across $C$ from $D$ (in the previous paragraph $n=2$ corresponds to one $e$- and one $m$-anyon) with which any excitation can be created. Therefore, in total there are
$N=2^n$ distinct excited states that run across $C$ from $D$. The lower bound of the non-local entanglement entropy is then given by $\tilde{S}=n\log 2$, which coincides with the value of topological entanglement entropy of the ground state. 

One can attribute the coefficient $n$ to the number of distinct fundamental anyonic excitations that run across $C$ from $D$. Indeed, the quantity $\tilde{S}$ was used to analyze the number of distinct excitations in several models, such as excitations in fracton topological phases~\cite{PhysRevB.97.144106}, higher-form excitations in the 3D toric code (point or membrane type excitation)~\cite{kim2014}, and excitations in various quantum double with gapped boundary~\cite{bridgeman2021boundary}. In the following, we will claim that the non-local entanglement entropy $\tilde{S}$ is a useful quantity characterizing the distinct anyonic excitations subject to the geometric constraint.

\subsection{Entanglement entropy of the $\mathbb{Z}_2$ toric code on book-page lattice}
Now we are in a good stage to calculate the non-local entanglement entropy of the toric code on the book-page lattice. 
Let us consider two cases of four-partite system $ABCD$ in the book-page lattice separated by cylinder geometry as portrayed in geometries (I) and (II) in Fig.~\ref{cay0}. \par
We first concentrate on the case of (I). 
The height of the cylinder $CD$ and $B$ is given by $L$, meaning that individual $1D$ vertical line of the book-page lattice within the cylinder contains $L$-qubits. Also, we set the radius of the cylinder $CD$ and $B$ is characterized by generation $M$ and $N$ of the book-page lattice, respectively. Namely, in the horizontal direction, $CD/B$ contains qubits up to $M/N$ generation. 
Let us calculate the entanglement entropy of subsystem $CD$. 
An example of such cylinder geometry is shown in Fig.~\ref{ee33}.
Deferring the details of the calculations to Appendix.~\ref{app1}, one finds
\begin{equation}
    S_{CD}=\text{Area}(CD)-\log 2, 
\end{equation}
where the first term is the area term, Area$(CD)=N_{\partial_{CD}}\log 2$, where $N_{\partial_{CD}}$ is the number of vertex operators $A_v$ that cross the subsystem $CD$ and its complement. 
The second term represents the universal topological entanglement entropy. One can also find the entanglement entropy in other cylinder geometry as
\begin{equation}
     S_{BC}=\text{Area}(BC)-\log2,\;\; S_{B}=\text{Area}(B)-2(k-1)^{M}\log 2,\;\; S_{D}=\text{Area}(D)-2\log2. \label{four}
\end{equation}
Therefore, the non-local entanglement entropy in the case of (I) of the four-partite system in Fig.~\ref{cay0} gives\footnote{Note that the area terms are suppressed as $\text{Area}(BC)+\text{Area}(CD)=\text{Area}(B)+\text{Area}(D)$.}
\begin{equation}
    \tilde{S}_{(I)}=2(k-1)^{M}\log2.\label{re}
\end{equation}

A similar line of thoughts shows that the non0local entanglement entropy in the case of (II) in Fig.~\ref{cay0} is yields the same value as \eqref{re}:
\begin{equation}
    \tilde{S}_{(II)}=2(k-1)^{M}\log2.\label{re2}
\end{equation}

\subsection{The number of excitations}\label{m5}
\begin{figure}
    \begin{center}
    \begin{subfigure}[h]{0.10\textwidth}
  \includegraphics[width=1.5\textwidth]{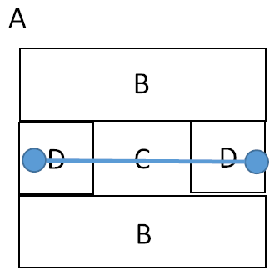}
     \caption{}\label{ee}
           \end{subfigure}
              \hspace{10mm}
            \begin{subfigure}[h]{0.5\textwidth}
    \includegraphics[width=1.0\textwidth]{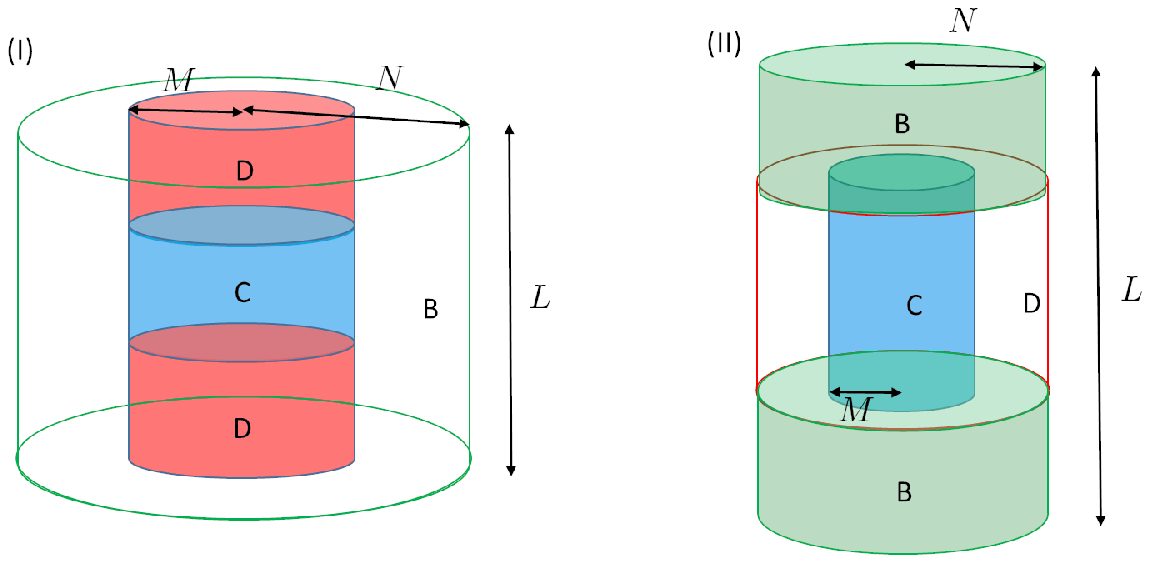}
         \caption{}\label{cay0}
             \end{subfigure}
             \hspace{10mm}
               \begin{subfigure}[h]{0.40\textwidth}
    \includegraphics[width=0.85\textwidth]{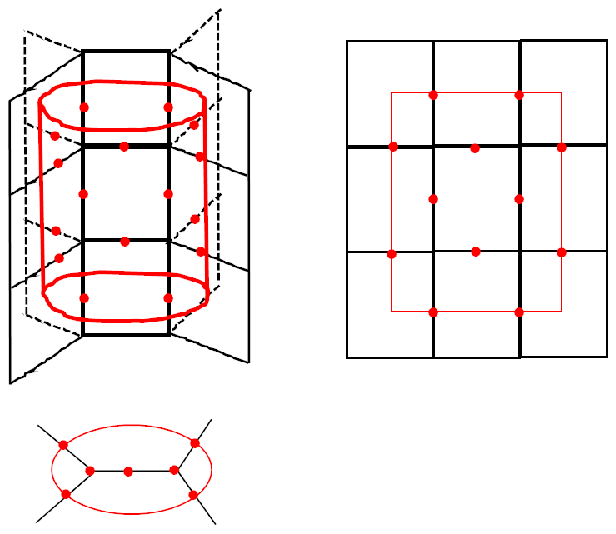}
         \caption{}\label{ee33}
             \end{subfigure}
   \end{center}
 \caption{(a) Four disjoint region $ABCD$ of the $\mathbb{Z}_2
 $ toric code on a 2D plane ($\rho_{ABCD}$ is a pure state), where anyonic excitation is running across $C$ from spatially separated $D$ where the end points of the excitation are located.
 (b) Two geometries of the four-partite system $ABCD$ in the book-page lattice model, where $A$ is the complement of $BCD$, which is not shown for simplicity. Each subsystem is separated by a cylinder geometry. (c) Left: Example of the cylinder geometry $CD$ (red line) of (I) in the previous figure with $k=3$, $L=3$, and $M=1$. The qubits within the cylinder $CD$ are marked by red dots. The top view of the same configuration is shown in the bottom. Right: The side view of the same geometry.  }
 \end{figure}
The non-local entanglement entropy conveys a clear physical meaning: the number of distinct excitations. The coefficient of \eqref{re} and \eqref{re2} properly counts the number of distinct anyons running across $C$ from $D$. 
Furthermore, even though the non-local entanglement entropy is identical in both cases of (I) and (II), the way it characterizes the excitations is rather different. \par
\begin{figure}
    \begin{center}
                      \begin{subfigure}[h]{0.35\textwidth}
    \includegraphics[width=1.0\textwidth]{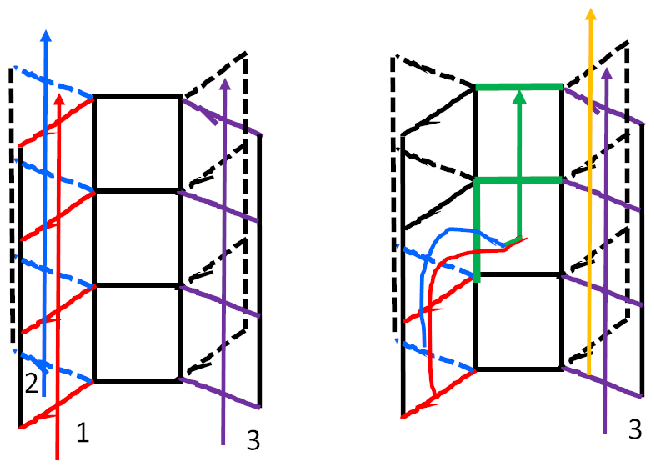}
         \caption{}\label{cay1}
             \end{subfigure}
             \hspace{10mm}
              \begin{subfigure}[h]{0.35\textwidth}
    \includegraphics[width=1.0\textwidth]{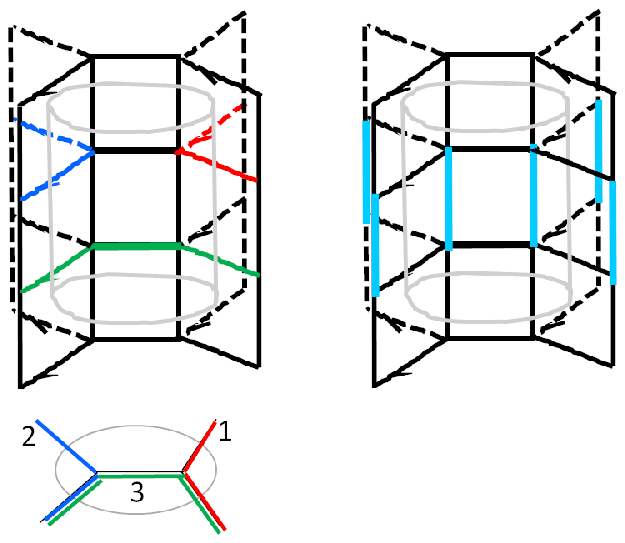}
         \caption{}\label{cayd}
             \end{subfigure}
                                   \begin{subfigure}[h]{0.30\textwidth}
    \includegraphics[width=1.0\textwidth]{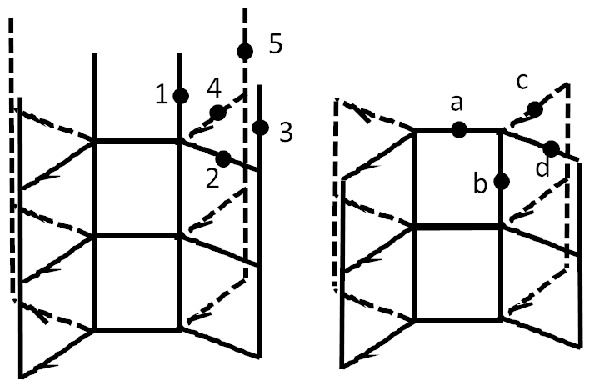}
         \caption{}\label{bdy1}
             \end{subfigure}
             \hspace{10mm}
              \begin{subfigure}[h]{0.20\textwidth}
    \includegraphics[width=1.0\textwidth]{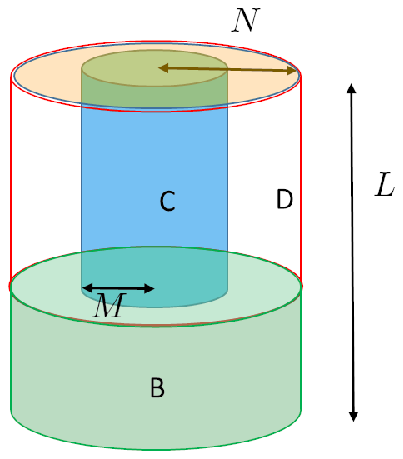}
         \caption{}\label{bdy2}
             \end{subfigure}
 \end{center}
       
      \caption{ (a) Three distinct $m$-anyons going across $C$ of case (I) in Fig.~\ref{cay0} with $k=3$. (left) The trajectories of these anyons are marked by red, blue and purple arrows. (right) Two $m$-anyons going along the red and blue arrow can be converted into $m$-anyon which goes along the green arrow. Likewise, $m$-anyon going along the yellow arrow is generated by combining $m$-anyon going along the purple arrow and the one along the green arrow. (b) Distinct trajectories of $e$- and $m$-anyons going across $C$ (grey line) of case (II) in Fig.~\ref{cay0} with $k=3$. (left) Three trajectories of $e$-anyon which go across $C$, marked by red, blue and green lines. The top vies of these trajectories are also shown below. (right) One trajectory of $m$-anyon going across $C$ marked by light blue links.
      (c) Rough/smooth boundary of the book-page lattice. (d) Four disjoint subsystem $ABCD$ (A, which is the complement of $BCD$ is not shown) with boundary condition (either rough or smooth boundary) on the top. }
   \end{figure}

Let us focus on the case with $M=1$ where $\tilde{S}=2(k-1)\log2$. (Practically, topological part of the entanglement entropy, like $\tilde{S}$, should be discussed in such a way that the size of the subsystems is longer than correlation length of the excitations. However, since the model has zero correlation length, one can discuss the topological properties of the entanglement entropy even in a small system size like the present case. )
There are $2(k-1)-1$ distinct $m$-anyons that go across $C$ from $D$. 
In Fig.~\ref{cay1}, with $k=3$, there are three distinct $m$-anyons whose trajectory is depicted by the red, blue, and purple arrows. The crucial point is that these three $m$-anyons are generators of other $m$-anyons whose trajectories run cross $C$. Indeed, $m$-anyon running along the green arrow in Fig.~\ref{cay1} is generated by two $m$-anyons marked by red and blue colors. Likewise, $m$-anyon going along the yellow arrow in Fig.~\ref{cay1} is generated by the ones of purple arrow and green arrow which is constructed by the two $m$-anyons with blue and red arrows. Thus, there are three $m$-anyons in total that exhaust all the magnetic excitations going across $C$ in the vertical direction. \par
The similar line of argument leads to that there are $2(k-1)^{M-1}-1$ distinct $m$-anyons running across $C$ vertically, by successively use of the fact that each vertical link, $k$ $m$-anyons are created from vacuum. 
%
There is only one distinct $e$-anyon going vertically in any case of $k$ and $M$, as it is deformable without any geometric constraint. In total, there are $2(k-1)^{M}-1+1=2(k-1)^M$ distinct anyons running through $C$ from $D$, which coincides with the coefficient of \eqref{re}.   \par
Now we turn to the four-partite system in the case (II) as portrayed in Fig.~\ref{cay0}. In this case $\tilde{S}$ characterizes the distinct excitations going across $C$ from $D$ in the \textit{horizontal} direction. In the case of $M=1$, there are three distinct excited states created by $e$-anyon. Indeed, we demonstrate three distinct paths of $e$-anyon that cross $C$ from $D$ in Fig.~\ref{cayd}. It is emphasized that even though $e$-anyon is the same as the toric code on the regular 2D plane, in the sense that a pair of $e$-anyons are created from vacuum, due to the non-trivial geometry of the bipartition between $C$ and $D$, there are non-trivial numbers of excited states of $e$-anyon, each of which contributes to $\tilde{S}$. On the contrary, as for $m$-anyon, there is only one excited state, forming like a membrane shape as depicted in the right geometry of Fig.~\ref{cayd} which goes across $C$. We also confirm that excited states $\sigma_{ABC}^{I} \;(1\leq I\leq 4)$ with $I$ corresponding to the three different configurations of $e$-anyon and one $m$-anyon, which are created by acting unitary operators on the ground state,
satisfies $\sigma_{ABC}^I\cdot \sigma_{ABC}^J=0\;(I\neq J)$ and that $I(A:C|B)|_{\sigma^{\prime}}=0$ with $\sigma^\prime=\frac{1}{4}\sum_I \sigma_{ABC}^I$, implying $\tilde{S}$ is saturated. 
For the generic values of $k$ and $M$, there are $2(k-1)^M-1$ distinct excited states coming from $e$-anyon, whereas there is only one excited state coming from $m$-anyon, thus, in total, there are $2(k-1)^{M}-1+1=2(k-1)^M$ distinct excitations running through $C$ from $D$, which again coincides with the coefficient of~\eqref{re2}.

\subsection{Consistency check -- boundary terms}
We further corroborate the counting argument of the distinct excitations which contribute to the non-local entanglement entropy $\tilde{S}$ by introducing the boundary terms. We impose the boundary condition on top of the book-page lattice in the analogy to the ones in the planner toric code~\cite{bravyi1998quantum}. 
\par
In the case of the rough boundary, plaquette terms $B_p$ at the boundary is incomplete in the sense that one link is missing.
Examples of such plaquette terms are given by $Z_1Z_2Z_3$ and $Z_1Z_4Z_5$ in Fig.~\ref{bdy1}. In the case of smooth boundary, the vertex terms $A_v$ is incomplete, missing one link. An example is given by $X_aX_bX_cX_d$ in Fig.~\ref{bdy1}.
We introduce an cylinder geometry portrayed in Fig.~\ref{bdy2} with its top being attached to the boundary of the book-page lattice. 
As we explain in the previous subsection, there are $2(k-1)^M-1$ distinct excitations of $e$-anyon and only one type of $m$-anyon contributing to $\tilde{S}$. Therefore, when the cylinder geometry has
the rough~(smooth) boundary condition on the top, the electronic~(magnetic) excitation is absorbed. 
As we leaned in the previous subsection, in geometry (II) given in Fig.~\ref{cay0}, there are $2(k-1)^M-1$ excited states coming from $e$-anyon and one excited state from $m$-anyon. In the presence of the boundary, either $e$- or $m$-anyon is absorbed, leading to that the non-local entanglement entropy is reduced to 
$\tilde{S}=\log2$ or $[2(k-1)^M-1]\log2$. In fact, calculations show that the non-local entanglement entropy with rough boundary reads

\begin{equation}
    \tilde{S}_r=\log2
\end{equation}
whereas the non-local entanglement entropy becomes in the presence of the smooth boundary
\begin{equation}
    \tilde{S}_s=[2(k-1)^{M-1}-1]\log2.
\end{equation}
These results coincide with the argument of the number of distinct $e$- and $m$-anyons that contributes to the non-local entanglement entropy;
in the case of the rough boundary, $2[(k-1)^M-1]$ excited states from $e$-anyon is absorbed whereas in the case of the smooth boundary, one excited state from $m$-anyon gets absorbed. 
\par

\section{Discussion and conclusion}\label{6}
In this work, we have highlighted the interplay between fractional excitations in $\mathbb{Z}_2$ topologically ordered phases and the geometric property of the Cayley tree, which is one of the examples of hyperbolic lattices by investigating two properties: the GSD counting on a closed surface and the non-local entanglement entropy. Both quantities characterize how the geometric structure of the Cayley tree affects the anyon excitations. \par
Let us make a brief comment on the generalization to other topologically ordered phases. We also study the $\mathbb{Z}_2$ modified surface code on the book-page lattice~\cite{surfacecode2012}. One distinction from the toric code is that one can design the model such that both of $e$- and $m$-anyons are subject to unusual fusion rules. Such a feature can clearly be seen in the form of the non-local entanglement entropy in the presence of the boundary.
We relegate the details of these arguments to Appendix.~\ref{app}.
Also, by introducing a topological defect where some of the pages of the book-page lattice are removed,  a single isolated Abelian anyon is bound at the dislocation. More thorough discussion on this defect and other types of defects such as twist defect will be discussed in a future project. \par

The hallmarks of this hyperbolic lattice are:  (i) Unusual fusion rule of the $m$-anyon, schematically described by $I\to \underbrace{m\otimes m\otimes \cdots \otimes m}_\textit{k}$ with $I$ and $m$'s represent the vacuum sector and $k$ $m$-anyonic excitations created by an $X$-operator acting on a single vertical link. 
As a consequence, 
iterative uses of this fusion rule at each vertical link lead to the delocalization of
the wave function of the $m$-anyon as the generation increases. (ii) Although there is only one type of $e$-anyon which is the same as the one of the planar toric code, 
in the horizontal direction, there are huge numbers of distinct paths for $e$-anyon to move, giving rise to a large number of the GSD or the non-local entanglement entropy. As we discussed in Sec.~\ref{sec stability}, some of the distinct paths differ by a small loop, thus the GSD is reduced by perturbation whose order equals the perimeter of the small loop. (iii) In the vertical direction, which is in the dual description of (ii), there are a large number of distinct $m$-anyons. 
Such $m$-anyons contribute to a large number of the GSD or the non-local entanglement entropy. \par
In addition to the previous discussion, there are some future directions that we would pursue.
First, it would be interesting to see whether one can have $K$-matrix description of Abelian topologically ordered phases on the book-page lattice. Note that the generalized K-matrix theory on the book-page lattice would be quite different from that on the plane, because of the inequivalence between electric and magnetic excitations. In the planar 2D toric code, we can introduce the K-matrix $K= 2 \sigma_x$ for the effective field theory description. In contrast, on the book-page lattice, there could be more geometric structures encoded in the K-matrix. There is an attempt~\cite{Fradkin2015} to construct discretized $K$-matrix description of the Chern-Simon theory on non-trivial discrete lattices, such as tetrahedron, where the information about the geometry is given in the form of couplings between gauge fields and the ones between gauge field and flux. To investigate whether it is possible to establish the generalized $K$-matrix description of the topologically ordered phases on the book-page lattice in the similar vein could be an useful approach.
Second, it is intriguing to explore non-Abelian topologically ordered phases on the book-page lattice. A simple step is to put pairs of twist defects in the book-page lattice and study the fusion rules with the defects~\cite{PhysRevLett.105.030403}. It could also be an important issue to see how the non-Abelian anyon is delocalized on the book-page lattice. Third, it is known that there is an intimate relation between gauge symmetry in a theory in $(d+1)$-dimension and the global symmetry in $d$-dimension holographically. For instance, the $e$-$m$ duality in the 2D toric code is closely related to the Kramers-Wannier duality in the transverse Ising model on its 1d boundary~\cite{PhysRevResearch.2.033317}. 
Studying the transverse Ising model on the 1D Cayley tree 
would help us to make better understanding of the bulk-edge correspondence. We will report our results elsewhere. 
\par

\section*{Acknowledgement}
We thank Hillel Aharony, Bishwarup Ash, Erez Berg, David Mross, Yuval Oreg, Ananda Roy, Ady Stern for discussion. We also thank Vijay B Shenoy for answering our questions on his work~\cite{manoj2021arboreal}, and David Mross for comments on the manuscript. 
This work is partly supported by Koshland postdoc fellowship (B.H.).
\bibliographystyle{ieeetr}

\bibliography{main}

\begin{thebibliography}{10}

\bibitem{tsui}
D.~C. Tsui, H.~L. Stormer, and A.~C. Gossard, ``Two-dimensional
  magnetotransport in the extreme quantum limit,'' {\em Phys. Rev. Lett.},
  vol.~48, pp.~1559--1562, May 1982.

\bibitem{wen1989chiral}
X.~G. Wen, F.~Wilczek, and A.~Zee, ``Chiral spin states and
  superconductivity,'' {\em Phys. Rev. B}, vol.~39, pp.~11413--11423, Jun 1989.

\bibitem{wen1990}
X.~G. Wen {\em International Journal of Modern Physics B}, vol.~04, no.~02,
  pp.~239--271, 1990.

\bibitem{Kitaev2003}
A.~Kitaev {\em Annals of Physics}, vol.~303, no.~1, pp.~2 -- 30, 2003.

\bibitem{Elitzur:1989nr}
S.~Elitzur, G.~W. Moore, A.~Schwimmer, and N.~Seiberg, ``{Remarks on the
  Canonical Quantization of the Chern-Simons-Witten Theory},'' {\em Nucl. Phys.
  B}, vol.~326, pp.~108--134, 1989.

\bibitem{leinaas1977theory}
J.~M. Leinaas and J.~Myrheim, ``On the theory of identical particles,'' {\em Il
  Nuovo Cimento B (1971-1996)}, vol.~37, no.~1, pp.~1--23, 1977.

\bibitem{Wilczek1982}
F.~Wilczek {\em Phys. Rev. Lett.}, vol.~49, pp.~957--959, 1982.

\bibitem{laughlin1983anomalous}
R.~B. Laughlin, ``Anomalous quantum hall effect: an incompressible quantum
  fluid with fractionally charged excitations,'' {\em Physical Review Letters},
  vol.~50, no.~18, p.~1395, 1983.

\bibitem{1982Jackiw}
S.~Deser, R.~Jackiw, and S.~Templeton, ``Three-dimensional massive gauge
  theories,'' {\em Phys. Rev. Lett.}, vol.~48, pp.~975--978, Apr 1982.

\bibitem{witten1989quantum}
E.~Witten, ``Quantum field theory and the jones polynomial,'' {\em
  Communications in Mathematical Physics}, vol.~121, no.~3, pp.~351--399, 1989.

\bibitem{dennis2002topological}
E.~Dennis, A.~Kitaev, A.~Landahl, and J.~Preskill, ``Topological quantum
  memory,'' {\em Journal of Mathematical Physics}, vol.~43, no.~9,
  pp.~4452--4505, 2002.

\bibitem{Nayak2008}
C.~Nayak, S.~H. Simon, A.~Stern, M.~Freedman, and S.~Das~Sarma {\em Rev. Mod.
  Phys.}, vol.~80, pp.~1083--1159, 2008.

\bibitem{dijkgraaf1990topological}
R.~Dijkgraaf and E.~Witten, ``Topological gauge theories and group
  cohomology,'' {\em Communications in Mathematical Physics}, vol.~129, no.~2,
  pp.~393--429, 1990.

\bibitem{spt2013}
X.~Chen, Z.-C. Gu, Z.-X. Liu, and X.-G. Wen, ``Symmetry protected topological
  orders and the group cohomology of their symmetry group,'' {\em Phys. Rev.
  B}, vol.~87, p.~155114, Apr 2013.

\bibitem{PhysRevB.86.115109}
M.~Levin and Z.-C. Gu, ``Braiding statistics approach to symmetry-protected
  topological phases,'' {\em Phys. Rev. B}, vol.~86, p.~115109, Sep 2012.

\bibitem{Kapustin2014}
A.~Kapustin and R.~Thorngren, ``Anomalous discrete symmetries in three
  dimensions and group cohomology,'' {\em Phys. Rev. Lett.}, vol.~112,
  p.~231602, Jun 2014.

\bibitem{fisher1961some}
M.~E. Fisher and J.~W. Essam, ``Some cluster size and percolation problems,''
  {\em Journal of Mathematical Physics}, vol.~2, no.~4, pp.~609--619, 1961.

\bibitem{kim2014}
I.~H. Kim and B.~J. Brown, ``Ground-state entanglement constrains low-energy
  excitations,'' {\em Phys. Rev. B}, vol.~92, p.~115139, Sep 2015.

\bibitem{PhysRevB.97.144106}
B.~Shi and Y.-M. Lu, ``Deciphering the nonlocal entanglement entropy of fracton
  topological orders,'' {\em Phys. Rev. B}, vol.~97, p.~144106, Apr 2018.

\bibitem{levinwen2006}
M.~Levin and X.-G. Wen, ``Detecting topological order in a ground state wave
  function,'' {\em Phys. Rev. Lett.}, vol.~96, p.~110405, Mar 2006.

\bibitem{preskillkitaev2006}
A.~Kitaev and J.~Preskill, ``Topological entanglement entropy,'' {\em Phys.
  Rev. Lett.}, vol.~96, p.~110404, Mar 2006.

\bibitem{freedman2002z2}
M.~H. Freedman, D.~A. Meyer, and F.~Luo, ``Z2-systolic freedom and quantum
  codes,'' {\em Mathematics of quantum computation, Chapman \& Hall/CRC},
  pp.~287--320, 2002.

\bibitem{zemor2009cayley}
G.~Z{\'e}mor, ``On cayley graphs, surface codes, and the limits of homological
  coding for quantum error correction,'' in {\em International Conference on
  Coding and Cryptology}, pp.~259--273, Springer, 2009.

\bibitem{manoj2021arboreal}
N.~Manoj and V.~B. Shenoy, ``Arboreal topological and fracton phases,'' {\em
  arXiv preprint arXiv:2109.04259}, 2021.

\bibitem{Han2019}
H.~Yan, ``Hyperbolic fracton model, subsystem symmetry, and holography,'' {\em
  Phys. Rev. B}, vol.~99, p.~155126, Apr 2019.

\bibitem{Breuckmann2017hyperboliccode}
N.~P. {Breuckmann}, C.~{Vuillot}, E.~{Campbell}, A.~{Krishna}, and B.~M.
  {Terhal}, ``{Hyperbolic and semi-hyperbolic surface codes for quantum
  storage},'' {\em Quantum Science and Technology}, vol.~2, p.~035007, Sept.
  2017.

\bibitem{maciejko2021hyperbolic}
J.~Maciejko and S.~Rayan, ``Hyperbolic band theory,'' {\em Science advances},
  vol.~7, no.~36, p.~eabe9170, 2021.

\bibitem{PastawskiYoshidaHarlowPreskill2015holographiccode}
F.~{Pastawski}, B.~{Yoshida}, D.~{Harlow}, and J.~{Preskill}, ``{Holographic
  quantum error-correcting codes: toy models for the bulk/boundary
  correspondence},'' {\em Journal of High Energy Physics}, vol.~2015, p.~149,
  June 2015.

\bibitem{chamon}
C.~Chamon, ``Quantum glassiness in strongly correlated clean systems: An
  example of topological overprotection,'' {\em Phys. Rev. Lett.}, vol.~94,
  p.~040402, Jan 2005.

\bibitem{Haah2011}
J.~Haah, ``Local stabilizer codes in three dimensions without string logical
  operators,'' {\em Phys. Rev. A}, vol.~83, p.~042330, Apr 2011.

\bibitem{Vijay}
S.~Vijay, J.~Haah, and L.~Fu, ``Fracton topological order, generalized lattice
  gauge theory, and duality,'' {\em Phys. Rev. B}, vol.~94, p.~235157, Dec
  2016.

\bibitem{Kitaev2006}
A.~Kitaev, ``{Anyons in an exactly solved model and beyond},'' {\em Annals of
  Physics}, vol.~321, pp.~2--111, jan 2006.

\bibitem{grover2012}
Y.~Zhang, T.~Grover, A.~Turner, M.~Oshikawa, and A.~Vishwanath, ``Quasiparticle
  statistics and braiding from ground-state entanglement,'' {\em Phys. Rev. B},
  vol.~85, p.~235151, Jun 2012.

\bibitem{Hamma2005}
A.~Hamma, R.~Ionicioiu, and P.~Zanardi, ``Bipartite entanglement and entropic
  boundary law in lattice spin systems,'' {\em Phys. Rev. A}, vol.~71,
  p.~022315, Feb 2005.

\bibitem{bridgeman2021boundary}
J.~C. Bridgeman, B.~J. Brown, and S.~J. Elman, ``Boundary topological
  entanglement entropy in two and three dimensions,'' {\em Communications in
  Mathematical Physics}, pp.~1--36, 2021.

\bibitem{bravyi1998quantum}
S.~B. Bravyi and A.~Y. Kitaev, ``Quantum codes on a lattice with boundary,''
  {\em arXiv preprint quant-ph/9811052}, 1998.

\bibitem{surfacecode2012}
A.~G. Fowler, M.~Mariantoni, J.~M. Martinis, and A.~N. Cleland, ``Surface
  codes: Towards practical large-scale quantum computation,'' {\em Phys. Rev.
  A}, vol.~86, p.~032324, Sep 2012.

\bibitem{Fradkin2015}
K.~Sun, K.~Kumar, and E.~Fradkin, ``Discretized abelian chern-simons gauge
  theory on arbitrary graphs,'' {\em Phys. Rev. B}, vol.~92, p.~115148, Sep
  2015.

\bibitem{PhysRevLett.105.030403}
H.~Bombin, ``Topological order with a twist: Ising anyons from an abelian
  model,'' {\em Phys. Rev. Lett.}, vol.~105, p.~030403, Jul 2010.

\bibitem{PhysRevResearch.2.033317}
W.~Ji, S.-H. Shao, and X.-G. Wen, ``Topological transition on the conformal
  manifold,'' {\em Phys. Rev. Research}, vol.~2, p.~033317, Aug 2020.

\bibitem{fattal2004entanglement}
D.~Fattal, T.~S. Cubitt, Y.~Yamamoto, S.~Bravyi, and I.~L. Chuang,
  ``Entanglement in the stabilizer formalism,'' {\em arXiv preprint
  quant-ph/0406168}, 2004.

\bibitem{Brown2013}
B.~J. Brown, S.~D. Bartlett, A.~C. Doherty, and S.~D. Barrett, ``Topological
  entanglement entropy with a twist,'' {\em Phys. Rev. Lett.}, vol.~111,
  p.~220402, Nov 2013.

\bibitem{wen2003}
X.-G. Wen, ``Quantum orders in an exact soluble model,'' {\em Phys. Rev.
  Lett.}, vol.~90, p.~016803, Jan 2003.

\end{thebibliography}
\appendix
\setcounter{equation}{0}
\renewcommand\thefigure{\thesection.\arabic{figure}} 
\setcounter{figure}{0}

\renewcommand{\theequation}{\thesection\arabic{equation}}
\section{Calculation of entanglement entropy}\label{app1}
\subsection{Stabilizer formalism}\label{di}
In this appendix, we present a way to calculate the entanglement entropy of the $\mathbb{Z}_2$ toric code on the book-page lattice, based on~\cite{fattal2004entanglement}. (See also Refs.~\cite{Brown2013,kim2014}.)\par
The basic idea behind \cite{fattal2004entanglement} to calculate entanglement entropy of subsystem $A$, $S_A$, is that for stabilizers $\{G_i\}$ that cross $A$ and its complement $\overline{A}$ (i.e., that acts on both of $A$ and $\overline{A}$), we introduce stabilizers, $\tilde{G}_i$, which have the local support of $G_i$ on $A$. In this appendix, we call such stabilizers as restricted stabilizers.
Among these restricted stabilizers $\tilde{G}_i$, one can construct ``canonical form" in such a way that $\{\tilde{G}_{2i-1},\tilde{G}_{2i}\}=0$ with other commutation relations being trivial. In other words, the canonical form is constructed such that each restricted stabilizer anti-commutes with only one restricted stabilizer while commuting with other ones. 
The entanglement entropy $S_A$ is given by $S_A=N\log 2$ where $N$ represents the number of pairs of the canonical form. \par
\begin{figure}
\begin{center}
\begin{subfigure}[h]{0.45\textwidth}
    \includegraphics[width=\textwidth]{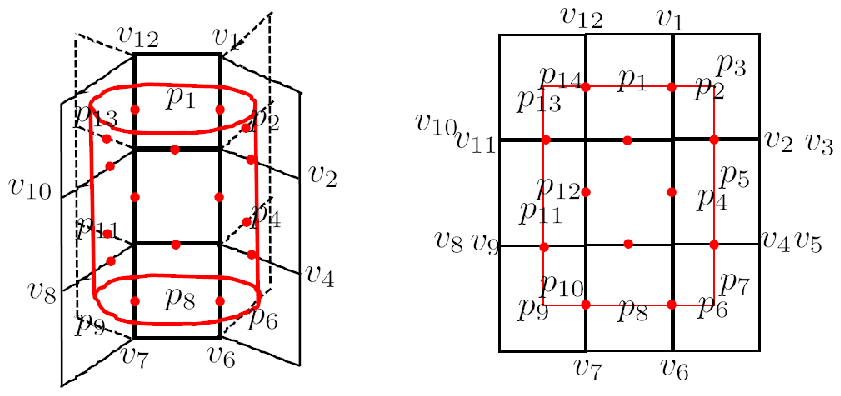}
         \caption{}
         \label{cylapp}
             \end{subfigure} 

  \hspace{5mm}
       
          \begin{subfigure}[h]{1.1\textwidth}
       \includegraphics[width=\textwidth]{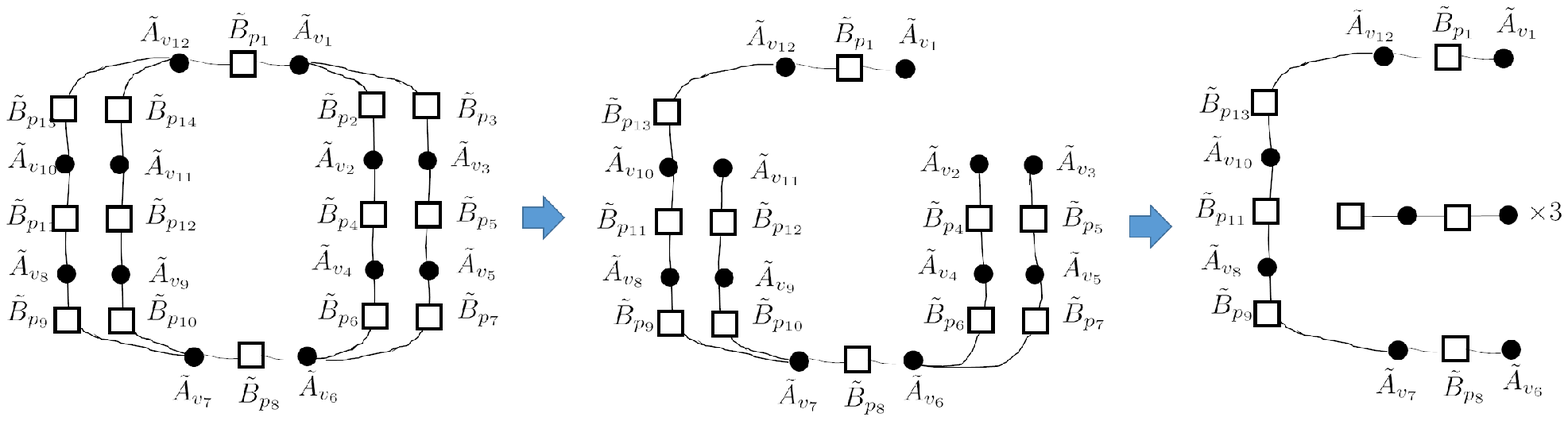}
         \caption{}\label{diagram}
          \end{subfigure}

      \caption{ (a) Subsystem of the book-page lattice separated by a cylinder geometry (red line). The qubits which are located within the cylinder are marked by red dots. Plaquette  and vertex terms that cross the cylinder and its complement are denoted by $p_i\;(1\leq  i\leq 14)$ and $v_j\;(1\leq j\leq 12)$ (some of index are not shown for the sake of clarity). The side view of this geometry is also exhibited on the right.
      (b) The diagram indicating anti-commutation relations between the restricted stabilizers. For restricted stabilizers which anti-commute with each other are connected by a line.}
        \label{diag}
        \end{center}
   \end{figure}

Let us apply this logic to our model. For simplicity, we focus on calculating the entanglement entropy of a  bipartite subsystem $CD$ separated by a cylinder geometry in the $\mathbb{Z}_2$ toric code on the book-page lattice with $k=3$. The diameter and height of the cylinder is set to be $M=1$ and $L=3$, as demonstrated in Fig.~\ref{cylapp}. There are 12 vertex terms $A_v$ and 14  plaquette terms $B_p$ that have nontrivial actions on both $CD$ and $\overline{CD}$. Accordingly, there are 12 and 14 restricted stabilizers $\tilde{A}_{v_i}$ and $\tilde{B}_{p_j}$. 

It is useful to introduce a diagram that represents the commutation relation between restricted stabilizers. The diagram is drawn as follows: the restricted stabilizer $\tilde{A_{v_1}}$ corresponding to the vertex term $A_{v_1}$ anti-commutes with three restricted stabilizers, $\tilde{B}_{p_2}$, $\tilde{B}_{p_1}$ and $\tilde{B}_{p_3}$, and commutes with other restricted stabilizers. 
Denoting black dots (white squares) as the restricted stabilizers of the vertex term (the three plaquette terms), we connect lines between the black dot and the white squares, indicating these restricted stabilizers anti-commute, giving the right corner of the diagram in the left of Fig.~\ref{diagram}. 
More precisely, the three lines between dots and squares represent the three relations
$\{\tilde{A}_{v_1},\tilde{B}_{p_s}\}=0\;(s=1,2,3)$. 
Anti-commutation relations between other restricted stabilizers are similarly discussed, yielding the reminder of the diagram portrayed in the left of Fig.~\ref{diagram}.\par
With this diagram, we construct canonical forms. To this end, we simplify the diagram. 
Noticing that there is a closed loops in the diagram, 
\begin{eqnarray*}
   \tilde{B}_{p_1}\to \tilde{A}_{v_1}\to \tilde{B}_{p_2}\to \tilde{A}_{v_2}\to \tilde{B}_{p_4}\to \tilde{A}_{v_4}\to \tilde{B}_{p_6}\to \tilde{A}_{v_6}\to\\\nonumber
   \tilde{B}_{p_8}\to \tilde{A}_{v_7}\to \tilde{B}_{p_{10}}\to \tilde{A}_{v_9}\to
 \tilde{B}_{p_{12}}\to \tilde{A}_{v_{11}}\to \tilde{B}_{p_{14}}\to\tilde{A}_{v_{12}}\to \tilde{B}_{p_1}
\end{eqnarray*}
one can redefine $\tilde{B}_{p_1}$ as
\begin{equation*}
    \tilde{B}_{p_1,new}=(\tilde{B}_{p_1}\tilde{B}_{p_2}\tilde{B}_{p_4}\tilde{B}_{p_6}\tilde{B}_{p_8}\tilde{B}_{p_{10}}\tilde{B}_{p_{14}})_{old},
\end{equation*}
   where we have introduced subscript ``new"
 and ``old" to emphasize that $\tilde{B}_{p_1}$ is redefined, 
 and verify that this newly defined $\tilde{B}_{p_1}$ commutes with any restricted stabilizer, allowing us to omit the white square corresponding to $\tilde{B}_{p_1}$. The similar procedure can be done when we find a closed loop in the diagram, and it is straightforward to show that one restricted stabilizer is omitted per closing loop. In the present case, 
two more squares, corresponding to $\tilde{B}_{p_2}$ and $\tilde{B}_{p_{14}}$ are dropped, giving the middle diagram of Fig.~\ref{diagram}.\par
In the middle of the diagram in Fig.~\ref{diagram}, focusing on the subdiagram along the line, $\tilde{A}_{v_6}\to \tilde{B}_{p_6}\to \tilde{A}_{v_4}\to \tilde{B}_{p_4}\to \tilde{A}_{v_2}$, we redefine the restricted stabilizers as
\begin{eqnarray}
    \tilde{A}_{v_6,new}&=&(\tilde{A}_{v_6}\tilde{A}_{v_4}\tilde{A}_{v_2})_{old}\nonumber\\
     \tilde{A}_{v_4,new}&=&(\tilde{A}_{v_4}\tilde{A}_{v_2})_{old}.\label{aaa1}
\end{eqnarray}
With these redefined stabilizers, one can show that
\begin{align}
    \{\tilde{A}_{v_4},\tilde{B}_{p_{6}}\}=0,\ \{\tilde{A}_{v_2},\tilde{B}_{p_{4}}\}=0, \nonumber \\ [\tilde{A}_{v_4},\tilde{B}_{p_{4}}]=0, \  [\tilde{A}_{v_6},\tilde{B}_{p_{6/4}}]=0, \nonumber 
\end{align}
indicating the two pairs, $ \tilde{A}_{v_4},\tilde{B}_{p_{6}}$ and $ \tilde{A}_{v_2},\tilde{B}_{p_{4}}$ are in canonical form, which allows us to detach the line $\tilde{B}_{p_6}\to \tilde{A}_{v_4}\to \tilde{B}_{p_4}\to \tilde{A}_{v_2}$ from the diagram. the similar consideration and redefining stabilizers in the similar fashion as \eqref{aaa1} enables us to detach the subdiagrams along the line $\tilde{B}_{p_7}\to \tilde{A}_{v_5}\to \tilde{B}_{p_5}\to \tilde{A}_{v_3}$ and $\tilde{B}_{p_{10}}\to \tilde{A}_{v_9}\to \tilde{B}_{p_{11}}\to \tilde{A}_{v_{12}}$ and obtain four pairs of restricted stabilizers each of which is in canonical form, giving the right diagram in Fig.~\ref{diagram}. We are left with the diagram, which is in a line shape. Similarly to \eqref{aaa1}, redefining restricted stabilizers such that
\begin{eqnarray*}
      \tilde{A}_{v_1,new}&=&(\tilde{A}_{v_1}\tilde{A}_{v_{12}}\tilde{A}_{v_{10}}\tilde{A}_{v_8}\tilde{A}_{v_{7}}\tilde{A}_{v_{6}})_{old}, \\
      \tilde{A}_{v_{12},new}&=&(\tilde{A}_{v_{12}}\tilde{A}_{v_{10}}\tilde{A}_{v_8}\tilde{A}_{v_7}\tilde{A}_{v_6})_{old}\nonumber\\
     \tilde{A}_{v_{10},new}&=&(\tilde{A}_{v_{10}}\tilde{A}_{v_{8}}\tilde{A}_{v_{7}}\tilde{A}_{v_{6}})_{old},\\
     \tilde{A}_{v_{8},new}&=&(\tilde{A}_{v_{8}}\tilde{A}_{v_{7}}\tilde{A}_{v_{6}})_{old},\\
     \tilde{A}_{v_{7},new}&=&(\tilde{A}_{v_{7}}\tilde{A}_{v_{6}})_{old}, 
\end{eqnarray*}
one can verify that five pairs $(\tilde{A}_{v_6}, \tilde{B}_{p_6})$, $(\tilde{A}_{v_7}, \tilde{B}_{p_9})$, $(\tilde{A}_{v_8}, \tilde{B}_{p_{11}})$, $(\tilde{A}_{v_10}, \tilde{B}_{p_{13}})$, and $(\tilde{A}_{v_{12}}, \tilde{B}_{p_1})$ are in canonical form. Together with six pairs that we have obtained previously, there are 11 pairs of restricted stabilizers which are in canonical form. Therefore, the entanglement entropy $S_{CD}$ is given by
\begin{equation*}
    S_{CD}=11\log 2.
\end{equation*}
Recalling the number of vertex terms that cross $CD$ and $\overline{CD}$ is 12, one can rewrite this result as 
$S_{CD}=12\log2-\log2$. The first term denotes the ``area" term, corresponding to the number of vertex terms that have nontrivial actions on both $CD$ and $\overline{CD}$, whereas the second one denotes topological entanglement entropy~\cite{levinwen2006,preskillkitaev2006}, hallmark of the topologically ordered phases. \par
By the same token, we can obtain the entanglement entropy of a cylindrical subsystem $CD$ with its center being located along the spine of the book-page lattice
for generic values of $k$, $L$ and $M$ :
\begin{equation}
    S_{CD}=N_{\partial_{CD}}\log2-\log2, 
\end{equation}
where the first term corresponds to the area term, reading $N_{\partial_{CD}}=4+2(k-1)^M(L-1)$.
Evaluation of the entanglement entropy for other geometries, $BC$, $B$, and $D$ can be similarly discussed, leading to \eqref{four}.

   \subsection{Alternative approach}\label{m4}
There is an alternative approach to calculate the entanglement entropy given  in~\cite{Hamma2005}, which we briefly review here. We will resort to this approach in the later section. 
Let $\mathcal{H}$ be the Hilbert space given by $\mathcal{H}=\mathcal{H}_{1/2}^{\otimes n}$, where $\mathcal{H}_{1/2}=\text{span}\{\ket{0},\ket{1}\}$, \textit{i.e.,} tensor product of spin-$1/2$ states ($\ket{0}$/$\ket{1}$ corresponds to spin-up/-down state in the spin-$z$ basis).
We are interested in stabilized states $\ket{\psi}\in\mathcal{H}$ such that $U_s\ket{\psi}=\ket{\psi}$ for a given set of $\mathbb{Z}_2$ stabilizers $\{U_s\}$ ($U_s^2=1$), which are mutually commuting operators acting on $\mathcal{H}$. 
Introducing $G=S_{1/2}^{\otimes n}$ with $S_{1/2}=\{\mathbb{I},X\}$ ($X$: the Pauli matrix), we define the ground state as 
\begin{equation}
    \ket{\psi_0}=\frac{1}{\sqrt{2^{N_0}}}\sum_{g\in G}g\ket{0}.
\end{equation}
Here, $\ket{0}=\ket{0}^{\otimes n}$ and $N_0 = |G|$. Obviously, this state is the stabilized state.  \par
The density matrix of the ground states has the form $\rho=\frac{1}{2^{N_0}}\sum_{g,g^{\prime}\in G}g\ket{0}\bra{0}gg^{\prime}$. 
In a bipartite system $AB$, factorizing the stabilizers $g$ $(g^\prime)$ as $g_A\otimes g_B ~(g^\prime_A\otimes g^\prime_B)$ which acts on the factorized Hilbert space $\mathcal{H}_A\otimes\mathcal{H}_B$, the reduced density matrix in $A$ is calculated to be~\cite{Hamma2005}
\begin{eqnarray}
     \rho_A&=&\text{Tr}_B({\rho})=
     \text{Tr}_B\Bigl[\frac{1}{2^{N_0}}\sum_{\substack{g,g^{\prime}\in g_A\otimes g_B
        }}
        g_A\ket{0_A}\bra{0_A}g_Ag{^\prime}_A
        \otimes  g_B\ket{0_B}\bra{0_B}g_Bg{^\prime}_B\Bigr]\nonumber\\
       &=&   
     \frac{1}{2^{N_0}}\sum_{\substack{g\in G\\
       g^{\prime}\in G_A
        }}
        g_A\ket{0_A}\bra{0_A}g_Ag^{\prime}_A\label{ee55},
\end{eqnarray}
where $G_{A/B}$ represents a set of stabilizers containing $X$ operators that act only on $A/B$. 
From the second to the last equation, one obtains the constraint $g^{\prime}_B=\mathbb{I}_B$ when tracing over $B$, which forces $g^\prime$ to be the form $g^{\prime}=g_A\otimes g_B\in G_A$, arriving at the last equation. In the last equation in \eqref{ee55}, we sum over $g\in G=g_A\otimes g_B$. Since $B$ is already traced out, one rewrite this summation via $\sum_{\substack{g\in G}}\to 2^{d_B}\sum_{\substack{g\in G/G_B}} $ with $d_B$ being the number of stabilizers that acts withing $B$, allowing us to rewrite \eqref{ee55} as
\begin{equation}
   \rho_A=\frac{2^{d_B}}{2^{N_0}}\sum_{\substack{g\in G/G_B\\
       g^{\prime}\in G_A
        }}
        g_A\ket{0_A}\bra{0_A}g_Ag^{\prime}_A. \label{ee3}
\end{equation}
To get the entanglement entropy, square~\eqref{ee3} to find [$d_A$: the number of stabilizers that act within $A$]
\begin{equation}
    \rho_A^2=\frac{2^{d_A+d_B}}{2^{N_0}}\rho_A,
\end{equation}
from which the entanglement entropy is given by~\cite{Hamma2005}
\begin{equation}
    S(\rho_A)=-\frac{\partial}{\partial n}\text{Tr}(\rho_A^n)\bigg|_{n=1}=(N_0-d_A-d_B)\log2.\label{fomee}
\end{equation}
Applying to this formula to our model yields \eqref{four}, as it should be. 

\begin{figure}
\begin{center}
  \begin{subfigure}[h]{0.35\textwidth}
    \includegraphics[width=\textwidth]{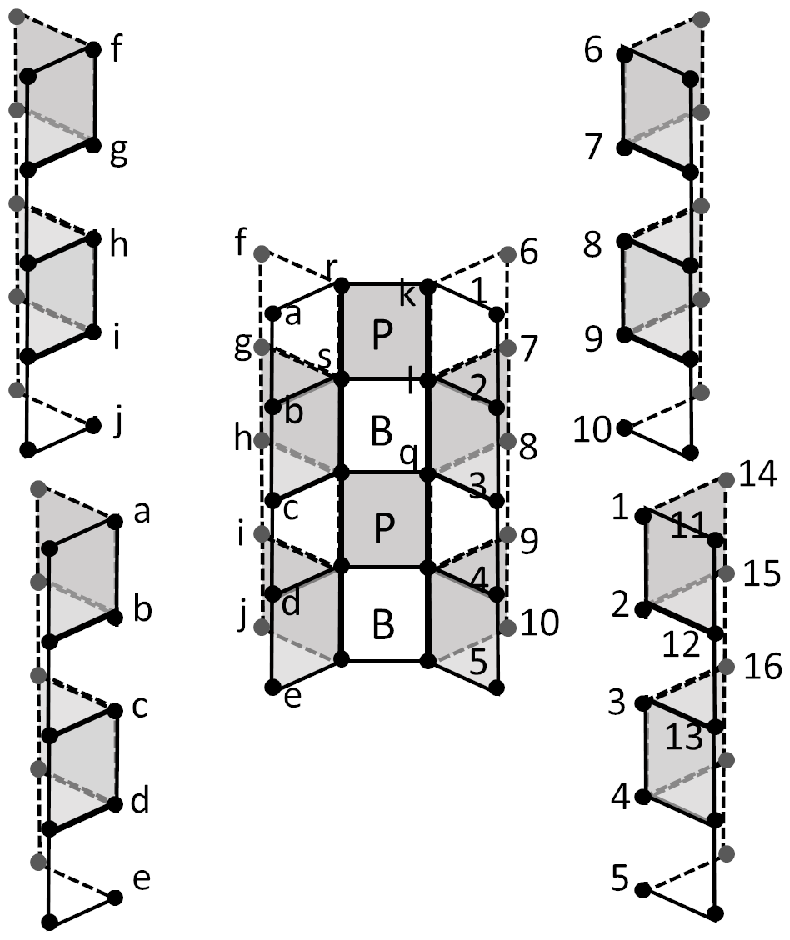}
         \caption{}\label{bpa}
  \end{subfigure} 
  \begin{subfigure}[h]{0.2\textwidth}
    \includegraphics[width=\textwidth]{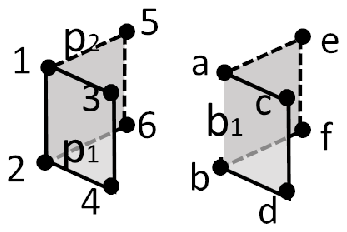}
         \caption{}\label{bandp}
  \end{subfigure} 
\begin{subfigure}[h]{0.2\textwidth}
       \includegraphics[width=\textwidth]{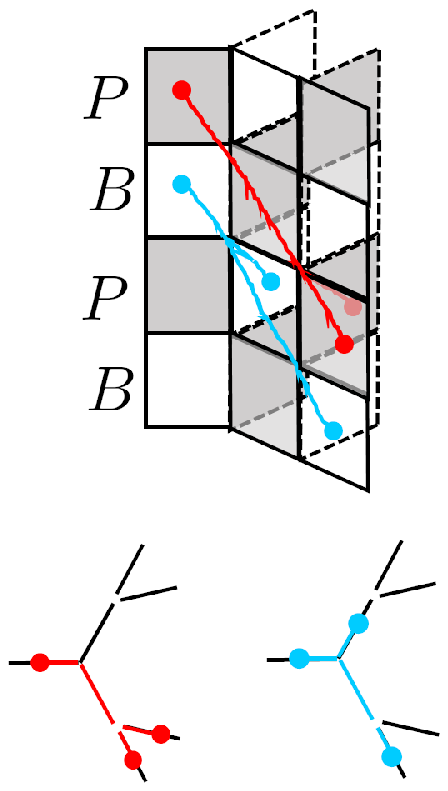}
         \caption{}
         \label{ex}
 
          \end{subfigure}
 \caption{(a) Modified $\mathbb{Z}_2$ surface code on the book-page lattice with $k=3$ up to the first generation. Two colors, white and grey are introduced marked in a zigzag pattern. Furthermore, we introduce two types of columns $P$ and $B$ in an alternating pattern. (b) Examples of two $X$-plaquette terms (left) and a $X$-book term (right). Examples of trajectories of $e$-($m$-) anyons which violate $X$- ($Z$-) plaquette or book    terms, marked by red (light blue) line with its top view being demonstrated below. }
       \end{center}
   \end{figure}
\section{Details of the $\mathbb{Z}_2$ surface code on the book-page lattice}\label{app}

We consider putting a different $\mathbb{Z}_2$ topologically ordered phase on
the book-page lattice, namely, $\mathbb{Z}_2$ surface code~\cite{wen2003,surfacecode2012} on the book-page lattice.  
A distinction from the case of the toric code is that one can design the model in such a way that $e$-anyon is also subject to the unusual fusion rule as well as the $m$-anyons. 
\subsection{Model}
Let us start with the construction of the lattice, which quite resembles the one discussed in Sec.~\ref{m22} except the fact that qubits are located at each node, not link. An example is shown in Fig,~\ref{bpa}.
To define Hamiltonian, we introduce four types of terms, $X/Z$-plaquette terms and $X/Z$-book terms. Each $X/Z$-plaquette term is defined by multiplication of four $X/Z$ operators at the corner of a plaquette. In the case of $k=3$, examples of X-plaquette terms are $O_{p_1}^X=X_1X_2X_3X_4$ and $O_{p_2}^X=X_1X_2X_5X_6$ in Fig.~\ref{bandp}. In a subsystem consisting of $k-1$ plaquettes connected to a vertical link, the $X/Z$-book term is given by multiplication of $2k$ $X/Z$ operators. An example of a $X$-book term is $O_{b_1}^X=X_aX_bX_cX_dX_eX_f$ in Fig.~\ref{bandp}.  
We also introduce two types of column, $P$ and $B$ in alternating pattern along which plaquette terms and book terms are introduced.~\footnote{At the spine, there is no difference between plaquette and books terms. Both of them are defined by multiplication of four $X$ or $Z$ operators, depending on the grey or white color.} Furthermore, we mark the system by two colors, grey and white in zigzag pattern, analogously to the $\mathbb{Z}_2$ surface code on the 2D plane~\cite{wen2003,surfacecode2012}. According to the types of columns and colors, we introduce $X$- and $Z$- plaquette or book terms. In Fig.~\ref{bpa}, along the first column on the top (P), depending on the color, $Z$- or $X$-plaquette terms are introduced. Examples are $X_rX_SX_kX_l$, $Z_1Z_2Z_kZ_l$, $Z_6Z_7Z_kZ_l$, $X_1X_2X_{11}X_{12}$, and $X_1X_2X_{14}X_{15}$. Analogously, along the second column (B) in Fig.~\ref{bpa}, $X$- or $Z$-book terms are introduced. Examples are $X_lX_qX_2X_3X_7X_8$ and $Z_2Z_3Z_{12}Z_{13}Z_{14}Z_{15}$.

Hamiltonian is defined by
\begin{equation}
    H=-\sum_{p\in P_{g}}O_{p}^X-\sum_{p\in P_{w}}O_{p}^Z-\sum_{b\in B_{g}}O_{b}^X-\sum_{b\in B_{w}}O_{b}^Z.\label{z211}
\end{equation}
where sets of plaquettes with grey (white) color is represented by $P_g (P_w)$ and sets of subsystems, where book term is introduced, with grey (white) color by $B_g (B_w)$. Also, X- (Z-) plaquette term is described by $O_p^X (O_p^Z)$ as well as X- (Z-)book term is by $O_b^X (O_b^Z)$.
One can show each term in Hamiltonian~\eqref{z211} commute with each other. 
The ground states satisfy $O_{p/b}^X\ket{\phi}=O_{p/b}^{Z}\ket{\phi}=\ket{\phi}\;\;\forall p, b$.

As for excitations, similarly to the main text, anyons which are charged by $X$~($Z$)-plaquette or book terms are refereed to as $e$ ($m$)-anyons. 
As opposed to the case of the toric code, in the present case, the $e$- and $m$- anyon shows the Cayley tree pattern, as demonstrated in Fig.~\ref{ex}

   \subsection{Entanglement entropy}
\begin{figure}
    \begin{center}
          \begin{subfigure}[h]{0.09\textwidth}
       \includegraphics[width=\textwidth]{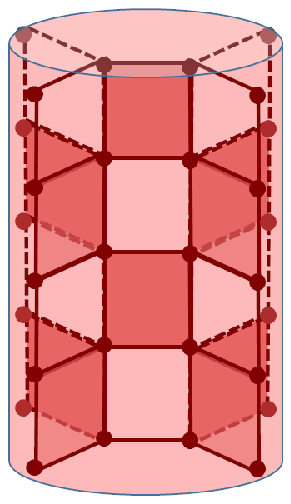}
         \caption{}\label{cy}
          \end{subfigure}
  \hspace{5mm}
     \begin{subfigure}[h]{0.49\textwidth}
    \includegraphics[width=\textwidth]{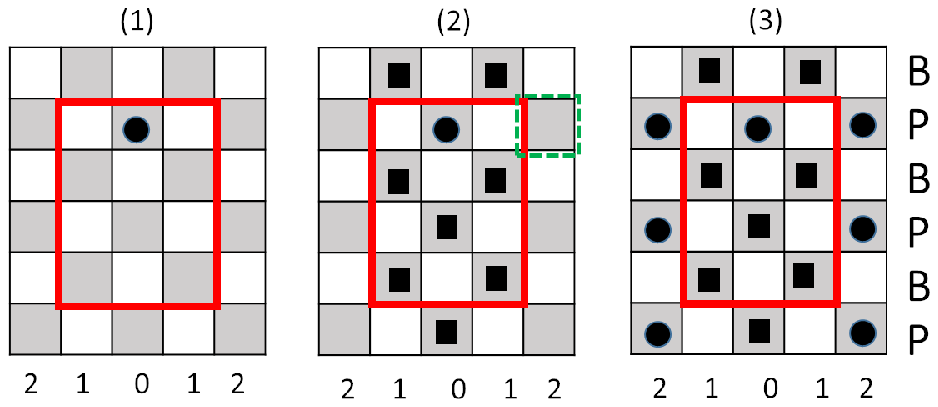}
         \caption{}\label{rule}
             \end{subfigure} 
 \end{center}
       
      \caption{ (a) Subsystem of the book-page lattice separated by a cylinder geometry with $k=3$. 
      (b) The side view of book-page lattice and the cylinder marked by red rectangle. 
      The numbers below each figure indicate the generation of the book-page lattice. The index ``B'' and ``P" stand for the two types of column that was explained above \eqref{z211}.
 }
        \label{diag}
   \end{figure}

 If the periodic boundary condition is imposed, some of the terms in~\eqref{z211} anti-commute.
Thus, the GSD counting in a closed surface, which relies on the stabilizer formalism, does not work to characterize the superselection sectors and geometric properties of the model in the present case. However, one can still argue the non-local entanglement entropy, which we now turn to in this subsection, properly characterizes the distinct anyonic excitations in the book-page lattice.
To calculate the non-local entanglement entropy, 
one could do the same trick as the one in Sec.~\ref{di} by introducing diagrams, which is now complected to draw. Instead of doing this, in this subsection, we try a different approach
outlined in Sec.~\ref{m4}~to calculate the entanglement entropy.
Let us consider a subsystem $A$ defined by qubits within a cylinder geometry. The height of the cylinder is $L$ qubits and its ``radius" is the first generation as shown in Fig.~\ref{cy}. For simplicity, we take $L$ to be an odd integer. However, the analysis works equally well for even $L$. We also portray the side view of this cylinder in Fig.~\ref{rule}.
Recalling the formula~\eqref{fomee} with replacing $B$ with the complement of $A$, i.e., $\overline{A}$, the entanglement entropy of the cylinder geometry $A$ consists of three terms: the total number of $X$-books and plaquette terms $N_0$, the number of $X$-books and plaquette terms that act within $A/\overline{A}$, $d_{A/\overline{A}}$. One has to carefully count $d_A/d_{\overline{A}}$ since depending on topology of the bipartite geometry, multiplication of $X$-books and plaquette terms may yield stabilizers that acts trivially on $A$ or $\overline{A}$.\par
Assuming the total number of plaquette and books terms with grey color in the entire system is given by $P_0+B_0$, we have $N_0=P_0+B_0$. The number of $X$-plaquette and book terms that act within $A$ is given by
\begin{equation}
    d_A=\frac{L-1}{2}+2\times\frac{L-1}{2},
\end{equation}
where the first term corresponds to the $X$-plaquette terms at spine and the second to books terms at the first generation. 
As for $d_{\overline{A}}$, which is the number of $X$-plaquette and book terms that act within $\overline{A}$, naively one would think it is given by subtraction the total number of the $X$-plaquette and book terms from $d_A$ and the ones that acts across $A$ and $\overline{A}$, namely, 
\begin{equation}
   d_{\overline{A}}\stackrel{?}{=} P_0+B_0-d_A-\frac{L+1}{2}\times 2(k-1)^2-1-2.\label{nm}
\end{equation}
However, \eqref{nm} is incorrect; it turns out that multiplication of the $X$-plaquette and book terms gives stabilizers that acts only on $\overline{A}$, which has to be also added to $d_{\overline{A}}$. 
Indeed, for subsets of plaquettes and books with grey color, $p\in \tilde{P_g}\subseteq P_g$ and $b\in \tilde{B}_g\subseteq B_g$, we have
\begin{equation}
    \prod_{p,b}O^X_{p}O^X_{b}=\mathbb{I}_A\otimes\label{ppap} G_{\overline{A}}
\end{equation} with $G_{\overline{A}}$ being non-trivial stabilizer acting on $\overline{A}$. To evaluate $d_{\overline{A}}$ properly, one has to count the number of ways of setting such subsets $p\in \tilde{P}_g$ and $b\in \tilde{B}_g$ to realize \eqref{ppap}.
Suppose such a multiplication contains one $X$-plaquette term at the spine within the cylinder (black dot in the first geometry of Fig.~\ref{rule}). Since the product [l.h.s of \eqref{ppap}] acts trivially on four qubits on each corner of the $X$-plaquette, four $X$-book terms in the first generation connected to the four qubits also have to be included in the product. Then other $X$-plaquette terms in the spine which share the same qubits as the four $X$-book terms have also to be included in the product. This line of thoughts can be iterated to find that once we assume the product contains one $X$-plaquette term at the spine within the cylinder, other $X$-plaquette terms at the spine and $X$-book terms at the first generation acting on qubits inside the cylinder have also to be included in the product in order for the product trivially acts on qubits between the spine and first generation inside the cylinder (corresponding to the black squares in the second geometry of Fig.~\ref{rule}). Furthermore, in order for the product acts trivially on qubits at the interface between the first and second generations inside the cylinder, $X$-plaquette terms at the second generation have to be included in the product. Focusing on $(k-1)^2$ $X$-plaquette terms on the right top of the second geometry of Fig.~\ref{rule} (green dashed line), there are $(k-1)(k-2)$ ways for the multiplication of $X$-plaquette terms to enter in the product. Other $X$-plaquette terms at the second generation can be similarly discussed. In total, there are $1+(L+1)(k-1)(k-2)$ ways to setting the product so that it satisfies \eqref{ppap}. 
When calculating $d_{\overline{A}}$, 
this number has to be added to \eqref{nm}, giving
\begin{equation}
    d_{\overline{A}}= P_0+B_0-d_A-\frac{L+1}{2}\times 2(k-1)^2-1-2+\bigl[1+(L+1)(k-1)(k-2)\bigr].
\end{equation}
We therefore arrive at
\begin{eqnarray}
    S_A&=&[N_{0}-d_A-d_{\overline{A}}]\log2=[2+(L+1)(k-1)]\log2\nonumber\\
    &=&[1+2+(L+1)(k-1)^2]\log 2-[1+(L+1)(k-1)(k-2)]\log 2.\label{b13}
\end{eqnarray}
In the last equation, we have decomposed the result into two. The first term corresponds to the area terms whereas the second term does to
the number of ways to setting the product of $X$-books and plaquette terms so that it acts trivially on $A$. Generally, this number indicates the topological property of excitations, independent of the local geometry of the system~\footnote{As a sanity check, when we set $k=2$, \eqref{b13} becomes $S_A=(L+4)\log2-\log2$, which is consistent with the form of the entanglement entropy of the $\mathbb{Z}_2$ surface code on the 2D plane~\cite{Brown2013}, implying that the second term is topological. }. Nevertheless, in the present case, the 
second term does depend on the height of the cylinder $L$ (and generation $M$ if one thinks about larger cylinder). As we will see below, such $L$ dependence is cancels out when calculating the non-local entanglement entropy, $\tilde{S}$.
The argument presented here is straightforwardly generalized to the case of any radius of the cylinder. 
\par
Now we are in a good place to study the non-local entanglement entropy of the $\mathbb{Z}_2$ surface code on the book-page lattice. 
\begin{figure}
    \begin{center}
          \begin{subfigure}[h]{0.49\textwidth}
       \includegraphics[width=\textwidth]{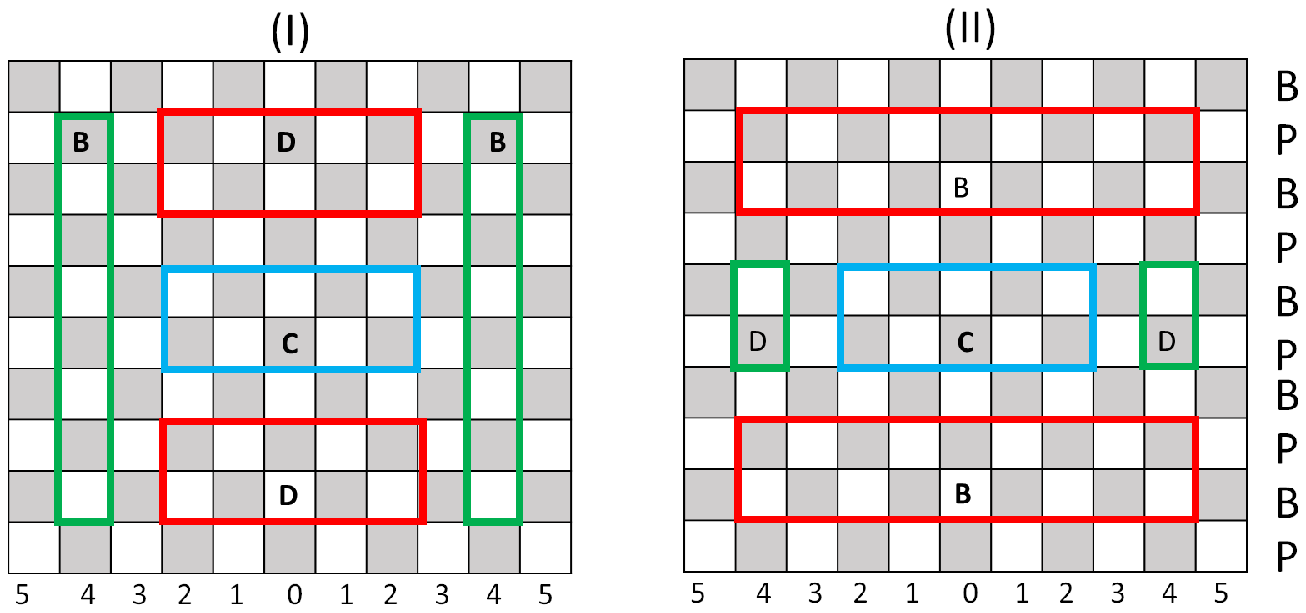}
         \caption{}\label{dg2}
          \end{subfigure}
           \begin{subfigure}[h]{0.24\textwidth}
       \includegraphics[width=\textwidth]{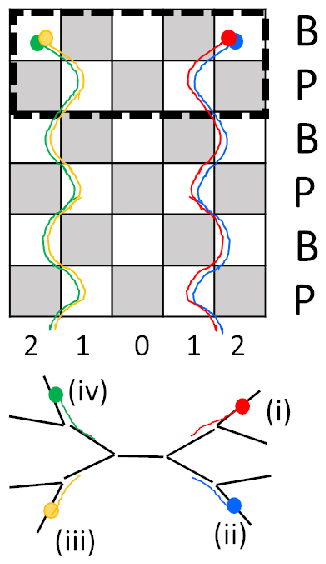}
         \caption{}\label{bpt4}
          \end{subfigure}
              \begin{subfigure}[h]{0.25\textwidth}
       \includegraphics[width=\textwidth]{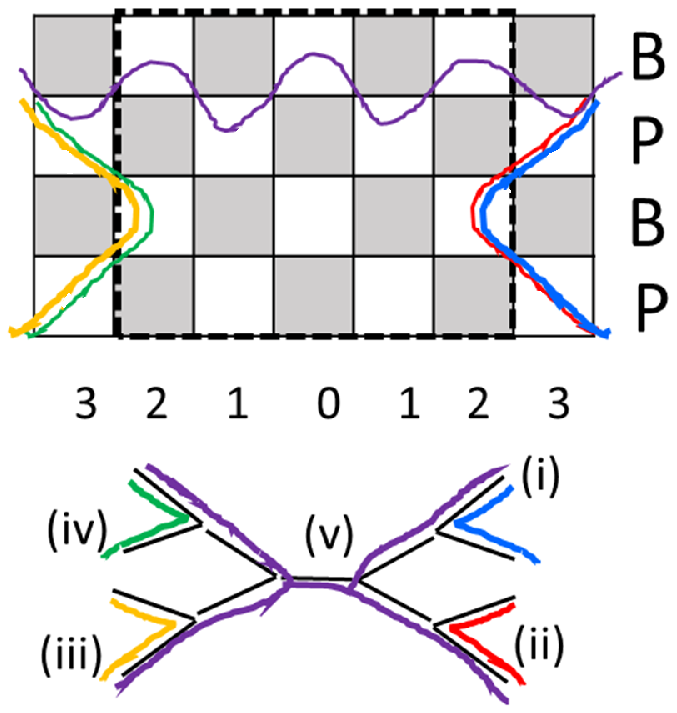}
         \caption{}\label{ppqq}
          \end{subfigure}
           \begin{subfigure}[h]{0.35\textwidth}
       \includegraphics[width=\textwidth]{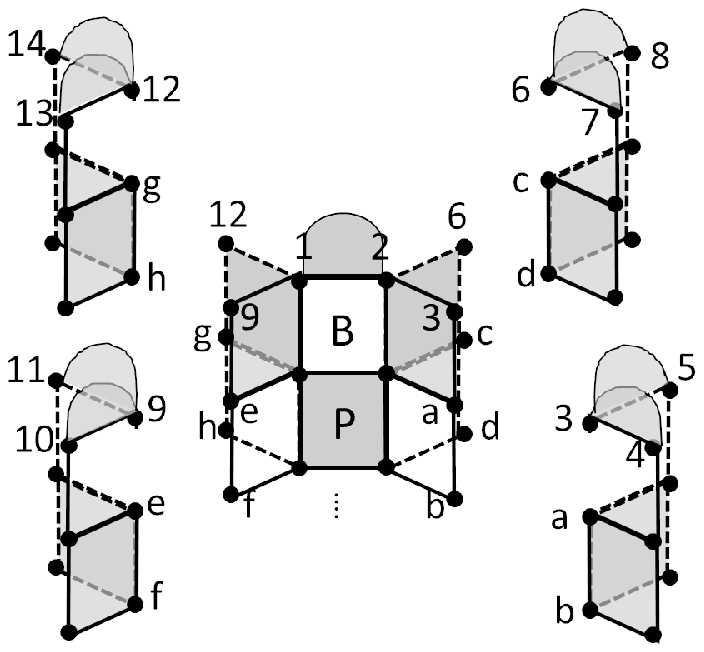}
         \caption{}\label{decorated2}
          \end{subfigure}
                \begin{subfigure}[h]{0.25\textwidth}
       \includegraphics[width=\textwidth]{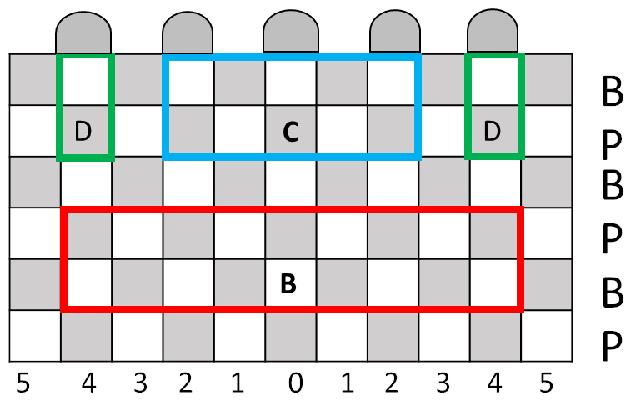}
         \caption{}\label{decorated}
          \end{subfigure}
 \end{center}
       
      \caption{ (a) Side view of four disjoint subsystems ($ABCD$, $A$ is not shown) in the book-page lattice. In these geometries, we set $M=2$.
      (b) Four distinct $m$-anyons that go across $C$ from $D$ (dashed line) in the geometry (I) in Fig.~\ref{bpt4}. Their trajectories are marked by red, blue, green, and yellow lines. The top view of these excitations are demonstrated below. (c) Five distinct paths for $m$-anyons going across $C$ (dashed line) from $D$ in the geometry (II) in Fig.~\ref{bpt4}. Their trajectories are marked by red, blue, green, purple, and yellow lines. Top view of these trajectories are also portrayed. 
      (d) Example of the decorated boundary ($X$-boundary) with $k=3$. The qubits with the same number or alphabet are identified. 
      (e) Side view of four disjoint subsystems ($ABCD$, $A$ is not shown) in the book-page lattice with decorated boundary on the top.  }
   \end{figure}
   Similarly to the main text, we envisage the same two geometries of four-partite system $ABCD$ whose side view is demonstrated in Fig.~\ref{dg2} ($A$ is complement of $BCD$). One can calculate $\tilde{S}$ defined in \eqref{stilde} following the similar logic presented in the previous paragraph. 
   For two geometries, (I) and (II) in Fig,~\ref{dg2}, $\tilde{S}$ is given by
  \begin{equation}
      \tilde{S}_{(I)}=\tilde{S}_{(II)}=2(k-1)^{M}\log2.\label{dth}
   \end{equation}
\subsection{Anyonic excitation interpretation}
We can interpret the result~\eqref{dth} as the number of distinct anyonic excitations that go along $C$ from $D$. Since the way we count such excitations closely parallels the one given in Sec.~\ref{m5}, we explain how to do it succinctly. Let us first focus on the non-local entanglement entropy in the geometry (I) in Fig.~\ref{dg2} with $M=2$. 
In the case of $k=3$, analogously to the discussion in Sec.~\ref{m5}, we can explicitly draw distinct three $m$-anyons going through $C$ from $D$ [(i), (ii), and (iii) in Fig.~\ref{bpt4}]. Other $m$-anyon excitation, like the one of (iv) in Fig.~\ref{bpt4} can be generated by combining the three $m$-anyons. 
One can also similarly discuss the number of distinct $e$-anyons. To summarize, denoting $D_{e/m}$ as the number of distinct $e/m$-anyons going through $C$ from D, the result is (including the generic cases of $k$ and $M$)
\begin{eqnarray*}
   D_e=
    \begin{cases}
    1+2(k-2)(k-1)+2(k-1)^3(k-2)+\cdots+ 2(k-1)^{M-1}(k-2)\;\;(M:\textbf{even}) \\
1+2(k-2)(k-1)+2(k-1)^3(k-2)+\cdots+ 2(k-1)^{M-2}(k-2) (M:\textbf{odd})
    \end{cases}\\
   D_m=
    \begin{cases}
    1+2(k-2)+2(k-1)^2(k-2)+\cdots+ 2(k-1)^{M-2}(k-2)\;\;(M:\textbf{even}) \\
    1+2(k-2)+2(k-1)^2(k-2)+\cdots+ 2(k-1)^{M-1}(k-2) (M:\textbf{odd})
    \end{cases}.
\end{eqnarray*}
In either case of $M$ being odd or even, $D_e+D_m=2(k-1)^M$ which is consistent with \eqref{dth}.
\par
Now we turn to $\tilde{S}$ in the geometry (II) in Fig.~\ref{dg2}. In this case, the number of distinct excitations amounts the number of distinct path for $e$-or $m$-anyons to cross $C$ from $D$. As an example, we demonstrate such distinct paths for $m$-anyons in Fig.~\ref{ppqq}. There are five distinct path for $m$-anyons with $k=3, M=2$ (other paths can be generated by these five paths.). Distinct path yield distinct excited states, each of which contributes to the non-local entanglement entropy. In the generic case of $k$ and $M$, we have
\begin{eqnarray}
 D_e=
    \begin{cases}
    1+2(k-2)+2(k-1)^2(k-2)+\cdots+ 2(k-1)^{M-2}(k-2)\;\;(M:\textbf{even}) \\
    1+2(k-2)+2(k-1)^2(k-2)+\cdots+ 2(k-1)^{M-1}(k-2) (M:\textbf{odd})
    \end{cases},\nonumber\\
   D_m=
    \begin{cases}
    1+2(k-2)(k-1)+2(k-1)^3(k-2)+\cdots+ 2(k-1)^{M-1}(k-2)\;\;(M:\textbf{even}) \\
1+2(k-2)(k-1)+2(k-1)^3(k-2)+\cdots+ 2(k-1)^{M-2}(k-2) (M:\textbf{odd}).
    \end{cases}
  \label{dedm}
\end{eqnarray}
which is again consistent with \eqref{dth} since $D_e+D_m=2(k-1)^M$. 
\par
One can also study the non-local entanglement entropy in the presence of decorated boundary. 
Analogously to the decorated boundary of the planner surface code~\cite{surfacecode2012}, we introduce boundary terms on the top of the book-page lattice. Examples of the $X$-boundary are shown in Fig.~\ref{decorated2} with $k=3$. Up to the second generation, we list such boundary terms: $X_1X_2$, $X_{3i}X_{3i+1}X_{3i+2}\;(1\leq i\leq 4)$. We can similarly consider the $Z$-boundary terms. Nice feature of these decorated boundary is that $X$- ($Z$-) boundaries absorb $m$- ($e$-) anyons, closely parallels the smooth and rough boundary of the toric code. \par
After introducing the decorated boundary, we envisage the four-partite system $ABCD$ with decorated $X$ boundary as shown in Fig.~\ref{decorated}, where $m$-anyons are condensed. Calculation shows 
 $\tilde{S}^{(II)}_{Xbdy}=D_e\log2$, where $D_e$ is given in \eqref{dedm}. This is consistent with the fact that the $X$ decorated boundary absorbs $m$-anyons, and only $e$-anyons can contribute to the non-local entanglement entropy. Similarly, one can study the non-local entanglement entropy with $Z$ decorated boundary to find $\tilde{S}^{(II)}_{Zbdy}=D_m\log2$ with $D_m$ is given in \eqref{dedm}.


\end{document}